\documentclass[fleqn,usenatbib]{mnras}

\usepackage{mathptmx}

\usepackage[T1]{fontenc}
\usepackage{ae,aecompl}

\usepackage{graphicx}	
\usepackage{amssymb}	
\usepackage{amsmath}

\title[Surface mass density ellipticity of clusters]{Measuring the surface mass density ellipticity of redMaPPer galaxy clusters using weak-lensing}

\author[Gonzalez et al.]{Elizabeth J. Gonzalez$^{1,2,3}$, Mart\'in Makler$^{1}$, Diego Garc\'ia Lambas$^{2,3}$,  \newauthor
Mart\'in Chalela$^{2,3}$, Maria E. S. Pereira$^{4}$, Ludovic Van Waerbeke$^{5}$, \newauthor
HuanYuan Shan$^{6,7}$, Thomas Erben$^{8}$ \\
$^{1}$ Centro Brasileiro de Pesquisas F\'{\i}sicas, Rio de Janeiro, RJ 22290-180, Brasil\\
$^{2}$ Instituto de Astronom\'{\i}a Te\'orica y Experimental (IATE-CONICET),
 Laprida 854, X5000BGR, C\'ordoba, Argentina.\\
$^{3}$ Observatorio Astron\'omico de C\'ordoba, Universidad Nacional de C\'ordoba, Laprida 854, X5000BGR, C\'ordoba, Argentina.\\
$^{4}$ Brandeis University, 415 South Street, Waltham, MA 02453, USA.\\
$^{5}$ Department of Physics and Astronomy, University of British Columbia, 6224 Agricultural road, V6T 1Z1 Vancouver, Canada. \\
$^{6}$Shanghai Astronomical Observatory (SHAO), Nandan Road 80, Shanghai 200030, China \\
$^{7}$University of Chinese Academy of Sciences, Beijing 100049, China \\
$^{8}$ Argelander-Institut f\"ur Astronomie, Auf dem H\"ugel 71, 53121 Bonn / Germany}

\begin{document}
\label{firstpage}
\pagerange{\pageref{firstpage}--\pageref{lastpage}}
\maketitle

\begin{abstract}
In this work we study the shape of the projected surface mass density distribution of galaxy clusters using weak-lensing stacking techniques. In particular, we constrain the average aligned component of the projected ellipticity, $\epsilon$, for a sample of redMaPPer clusters ($0.1 \leq z < 0.4$). We consider six different proxies for the cluster orientation and measure $\epsilon$ for three ranges of projected distances from the cluster centres. The mass distribution in the inner region (up to $700$\,kpc) is better traced by the cluster galaxies with a higher membership probability, while the outer region (from $700$\,kpc up to 5\,Mpc) is better traced by the inclusion of less probable galaxy cluster members. The fitted ellipticity in the inner region is $\epsilon = 0.21 \pm 0.04$, in agreement with previous estimates. We also study the relation between $\epsilon$ and the cluster mean redshift and richness. By splitting the sample in two redshift ranges according to the median redshift, we obtain larger $\epsilon$ values for clusters at higher redshifts, consistent with the expectation from simulations. In addition, we obtain higher ellipticity values in the outer region of clusters at low redshifts. We discuss several systematic effects that might affect the measured lensing ellipticities  and their relation to the derived ellipticity of the mass distribution.
\end{abstract}




\section{Introduction}
According to the current $\Lambda$CDM cosmological paradigm, structures in the Universe
are formed hierarchically from gravitational collapse due to initial density
fluctuations \citep{Kravtsov2012}. As matter collapses, gas condensation
results on star formation and eventual galaxy formation occur within these 
matter overdensities. Therefore, galaxies and galaxy systems are expected to 
reside on highly overdense dark matter clumps of increasing mass,
 named as dark matter halos. 
In this scenario, massive halos are formed from the accretion of smaller
ones. These low-mass halos are accreted in preferential directions, mainly
along the filamentary distribution. Therefore, 
dark matter halos are not expected to have spherical shapes. In fact, 
studies of dark matter halos in numerical simulations have 
shown that halo shapes can be well described by a triaxial model
with a tendency of being prolate \citep{Dubinski1991,Warren1992,Cole1996,Jing2002,Bailin2005,Hopkins2005,Kasun2005,Allgood2006,Paz2006,Bett2007,Munoz2011,Schneider2012,Despali2013,Velliscig2015,Vega-Ferrero2017}. 

Numerical simulations also predict that more massive halos tend to be
less spherical since they
are formed later, thus, they are less dynamically relaxed.
The same is expected for halos at larger redshifts \citep{Jing2002,Allgood2006,Munoz2011,Velliscig2015},
because they are affected by the direction of the last major merger and
the presence of filaments around them \citep{Bonamigo2015}.
Also, cluster of galaxies appear to be less spherical towards their
centres \citep{Warren1992,Jing2002,Bailin2005,Allgood2006,Schneider2012,Velliscig2015}.


Gravitational lensing provides a unique technique to
constrain the mean projected halo ellipticity.
A triaxial dark matter halo will produce
an azimuthal variation of the lensing
signal, causing a larger amplitude of the observed signal along
the projected major axis direction. This effect can be well modelled
to obtain the projected ellipticity for individual targets as long
as massive galaxy clusters or deep data are considered
\citep[e.g.][]{Oguri2010,Harvey2019,Okabe2020}. Nevertheless, when lower mass halos
are analysed, constraining
the lensing signal is a hard task. In the weak lensing regime, 
the combination of a large sample of
systems is needed to obtain a reliable determination of the halo 
ellipticity, known as stacking techniques \citep{Brainerd2000,Natarajan2000}. 
Another difficulty comes from the necessity of making
assumptions regarding the projected major-semi axis 
orientation in order to align the systems. Misalignment between 
the adopted angle and the true halo orientation would 
result in a depletion of the detected signal, thus underestimating
the halo ellipticity.  

In spite of these difficulties, several
studies were able to successfully measure the projected halo ellipticity
using weak lensing techniques for a wide range of masses. For
galaxy scale halos, assuming that the galaxy and the
halo are aligned, the average ratio between the aligned component of
the halo ellipticity and the ellipticity of the light distribution
could be successfully constrained \citep[][Schrabback in prep.]{Mandelbaum2006,Parker2007,Uitert2012,Schrabback2015}.
For the group- and cluster-scale, either the major semi axis
of the brightest cluster (or group) galaxy member (BCG),
or the distribution of the galaxy members, have been
used as proxies to trace the halo orientation. While
the BCG's major axis is the optimal proxy to trace the matter orientation
on small scales \citep[$\lesssim 250$\,kpc,][]{Uitert2017}, the
member distribution is better aligned with the matter at
the outskirts of the galaxy systems. Using this approach, the mean projected halo ellipticity
for these systems has been successfully estimated, obtaining an aligned component 
of $\sim 0.2 - 0.5$ \citep{Evans2009,Oguri2010,Clampitt2016,Uitert2017,Shin2018}.

In this work we present the analysis of the aligned component 
of the projected surface mass ellipticity for SDSS redMaPPer clusters \citep{Rykoff2014}
using weak-lensing stacking techniques. 
For this sake, we take advantage
of high quality public weak lensing surveys, combining these data to increase
the signal-to-noise of our measurements. In order to align the combined clusters, we estimate the surface density orientation angle of each galaxy cluster taking into account the satellite\footnote{Throughout this work we consider satellite galaxies as the galaxies that were classified as cluster members, besides the central galaxy. These galaxies are supposed to be located at the satellite halos, hosted in the main dark matter halo of the cluster.} distribution. We define different proxies to estimate this orientation, considering different weights and cuts for the satellite galaxies that are assumed to trace the mass distribution. Given the high-mass systems considered for the analysis and the good quality data used to obtain the lensing signal, we also study the relation between the derived projected ellipticity and the average cluster mass and redshift. Furthermore, we analyse the projected lensing signal at different ranges of distances from the cluster centres to obtain information about the orientation of the surface mass density distribution at the inner and outer parts of the clusters. The main motivation is to study the halo ellipticity as well as the orientation of the surface mass distribution at the outskirts of the cluster. 

This paper is organised as follow: In Sec. \ref{sec:data} we describe 
the data used in this work, the weak-lensing
catalogues and the sample of clusters. We describe how the projected matter orientation is
estimated for each cluster in Sec. \ref{sec:orientation}. In Sec. \ref{sec:analysis}
we describe the lensing analysis applied in order to derive the aligned component 
of the projected ellipticity. We present our results in Sec. \ref{sec:results}
and discuss different sources of biases in Sec. \ref{sec:bias}. Finally, we discuss our results  in
Sec. \ref{sec:discuss} and conclude in Sec. \ref{sec:conclusion}.  We adopt when necessary a standard cosmological model with $H_{0}$\,=\,$70h_{70}$\,km\,s$^{-1}$\,Mpc$^{-1}$, $ \Omega_{m} $\,=\,0.3, and $ \Omega_{\Lambda} $\,=\,0.7.

\section{Observational data}
\label{sec:data}

\subsection{Weak Lensing surveys}

In order to perform the weak lensing analysis we combine the catalogues from
four public weak-lensing surveys. All the combined data (except for KiDS-450) are based
on imaging surveys carried-out using the MegaCam 
camera \citep{Boulade2003} mounted on the Canada France Hawaii Telescope (CFHT), thus having similar
image quality. Data products were combined using \textsc{THELI} \citep{Erben2013}.
Moreover, all the source galaxy catalogues 
were obtained using \textit{lens}fit \citep{Miller2007,Kitching2008}
to compute the shape measurements and photometric redshifts are estimated
using the BPZ algorithm \citep{Benitez2000,Coe2006}.
For the analysis we applied the additive calibration correction factors
for the ellipticity components provided for each catalog. We also apply
a multiplicative shear calibration factor to the combined sample of galaxies
as suggested by \citet{Miller2013} (See subsection \ref{subsec:estimators}).
In the next subsections we will briefly describe the shear catalogues used.

From the weak lensing catalogs, we select the galaxies for the lensing study
by applying the following cuts to the \textit{lens}fit parameters: 
MASK $\leq$ 1, FITCLASS $= 0$ and $w > 0$.
Here MASK is a masking flag, FITCLASS is a flag parameter given by \textit{lens}fit 
which takes the value $0$ when the source is classified as a galaxy and
$w$ is a weight parameter that takes into account errors on the shape measurement
and the intrinsic shape noise, which ensures that galaxies have
well-measured shapes \citep[see details in ][]{Miller2013}. Background galaxies, defined as the galaxies that are located behind the galaxy clusters
and thus affected by the lensing effect, are selected
taking into account the photometric redshift information
with a similar criteria as the one used in previous studies
\citep[e.g.][]{Leauthaud2017,Pereira2018,Chalela2018,Gonzalez2019}.
We consider a galaxy as a background galaxy if it satisfies 
Z\_BEST$ > z_c + 0.1$ and ODDS\_BEST $> 0.5$, where
Z\_BEST is the photometric redshift estimated for each galaxy,
$z_c$ is the considered cluster redshift and 
ODDS\_BEST is a paramter that expresses the quality of 
Z\_BEST and takes values from 0 to 1. We neglect
the effect of the inclusion of foreground and/or 
cluster galaxies in the background sample, 
known as `boost factor', which cause a dilution effect in 
the lensing signal, since it is expected to be negligible
considering the cuts implemented in the background sample selection
\citep{Leauthaud2017,Shan2018,Blake2016}.
To assign background galaxies to each galaxy cluster
we use the public regular grid search algorithm \textsc{grispy}\footnote{\href{grispy}{https://github.com/mchalela/GriSPy}}
\citep{chalela2019grispy}. 
An analysis of the lensing signal computed for the individual catalogues is presented in Apppendix \ref{app:test} as a control test for their combination.

\subsubsection{CFHTLens}

The Canada-France-Hawaii Telescope Lensing Survey (CFHTLenS) weak lensing 
catalogs\footnote{CFHTLenS: http://www.cadc-ccda.hia-iha.nrc-cnrc.gc.
ca/en/community/CFHTLens}
are based on the data
collected as part of the CFHT Legacy Survey. This
is a multiband survey ($u^*g'r'i'z'$) that spans 154 square
degrees distributed in four separate patches W1, W2, W3 and W4 ($63.8$, $22.6$, $44.2$
and $23.3$ deg$^2$, respectively). The achieved limiting magnitude
is $i' \sim 25.5$ considering a 5$\sigma$ point source detection.
Further details regarding image reduction, 
shape measurements, and photometric redshifts can be found in
\citet{Hildebrandt2012,Heymans2012,Miller2013,Erben2013}. The shear catalog
is based on the $i-$band measurements, achieving a weighted galaxy source
density of $\sim 15.1$\,arcmin$^{-2}$.

\subsubsection{CS82}

The CS82 shear catalogue is based on the CFHT Stripe 82 survey, 
a joint Canada-France-Brazil project designed
with the goal of complementing existing SDSS Stripe 82 $ugriz$
photometry with high-quality $i-$band imaging suitable for weak
and strong lensing measurements \citep[e.g.,][]{Shan2014,Hand2015,Liu2015,Bundy2017,Leauthaud2017,Shan2017,Niemiec2017,Pereira2018}.
This survey spans over a window of $2 \times 85$\,deg$^2$, with an effective
area of $129.2$\,deg$^2$, after masking out bright stars and other
image artifacts. It has a median point spread function (PSF) of
$0.6''$ and a limiting magnitude $i' \sim 24$ \citep{Leauthaud2017}. 
The source galaxy catalogue has an effective weighted galaxy number density
of $\sim 12.3$\,arcmin$^{-2}$.

\subsubsection{RCSLens}

The RCSLens catalog\footnote{RCSLenS: https://www.cadc-ccda.hia-iha.nrc-cnrc.gc.ca/en/community/rcslens}
\citep{Hildebrandt2016} is based on the Red-sequence Cluster Survey 2
\citep[RCS-2,][]{Gilbank2011} a  multi-band  imaging  survey  in  the $griz-$bands with a depth 
of $\sim24.3$  in  the $r-$band, considering a point source at 7$\sigma$ detection level.
The survey spans over $\sim785$\,deg$^2$ distributed in 14 patches, the largest being
$10 \times 10$\,deg$^2$ and the smallest $6 \times 6$\,deg$^2$. The source catalogue 
based on the $r-$band imaging, achieves an effective weighted galaxy number density
of $\sim 5.5$\,arcmin$^{-2}$. A full systematic error analysis of the shear measurements
in combination with the photometric redshifts is presented in
\citet{Hildebrandt2016}, with additional error analyses of the
photometric redshift measurements presented in \citet{Choi2016}.

\subsubsection{KiDS-450}

The KiDS-450 catalog\footnote{KiDS-450:
http://kids.strw.leidenuniv.nl/cosmicshear2018.php}
\citep{Hildebrandt2017} is based on the third
data release of the Kilo Degree
Survey \citep[KiDS,][]{Kuijken2015} which spans over 447\,deg$^2$. 
This is a multi-band imaging survey ($ugri$)
carried out with the Omega-CAM CCD mosaic camera mounted on the VLT Survey Telescope (VST). 
Shear catalogues are based on the $r-$band images with a mean
PSF of $0.68''$ and a $5\sigma$ limiting magnitude of $25.0$. Shape measurements
are performed using an upgraded version of $lens$fit
algorithm \citep{Fenech2017}. 
The resultant source catalogue has  an effective weighted galaxy number density
of $\sim 8.53$\,arcmin$^{-2}$.

\subsection{redMaPPer clusters}

We use the redMaPPer public catalogue \citep[v6.3;][]{Rykoff2016}
which is based on a red-sequence algorithm for finding clusters \citep{Rykoff2014}, 
applied to the Sloan  Digital  Sky  Survey  Data  Release  8  photometric data
\citep{York2000,Aihara2011}, covering an
area of $10^4$ deg$^2$ , down to a limiting magnitude of $i = 21$ for galaxies. 
Briefly, the algorithm uses multi-band
colours to find overdensities of red-sequence galaxies around central galaxy candidates.
It computes the probability that each galaxy in the vicinity of the cluster is a red-sequence 
member, $p_{mem}$. This membership probability takes into 
account the galaxy colour, luminosity and projected distance
from the cluster centre.
The total richness of each identified cluster, $\lambda$, is computed 
as the sum of the membership probabilities over all of the galaxies within a scale-radius, $R_{\lambda}$,
considering luminosity- and radius-dependent weights.
$R_{\lambda}$ is related to the richness through \citep{Rykoff2014}:
\begin{equation}
    R_\lambda = 1\,\rm{Mpc}\, h^{-1} \left( \frac{\lambda}{100} \right)^{0.2}
\end{equation}
and is optimized together with the cluster richness, to maximize 
the signal-to-noise of the richness measurements. redMaPPer uses the 
$5-$band ($ugriz$) from SDSS data to self-calibrate the red-sequence, identify
red galaxy overdensities and to estimate the photometric redshift, $z_{\lambda}$,
for each cluster. It also assigns the centre of the halo  according to the position
of one of the brightest members (notice that it is not necessarily the brightest member)
taking into account each galaxy's luminosity, photometric redshift, and local galaxy density.
We consider the cluster redshift, $z_c$, as the spectroscopic redshift when it is
available, otherwise we use $z_{\lambda}$ instead.

\begin{figure}
    \centering
    \includegraphics[scale=0.7]{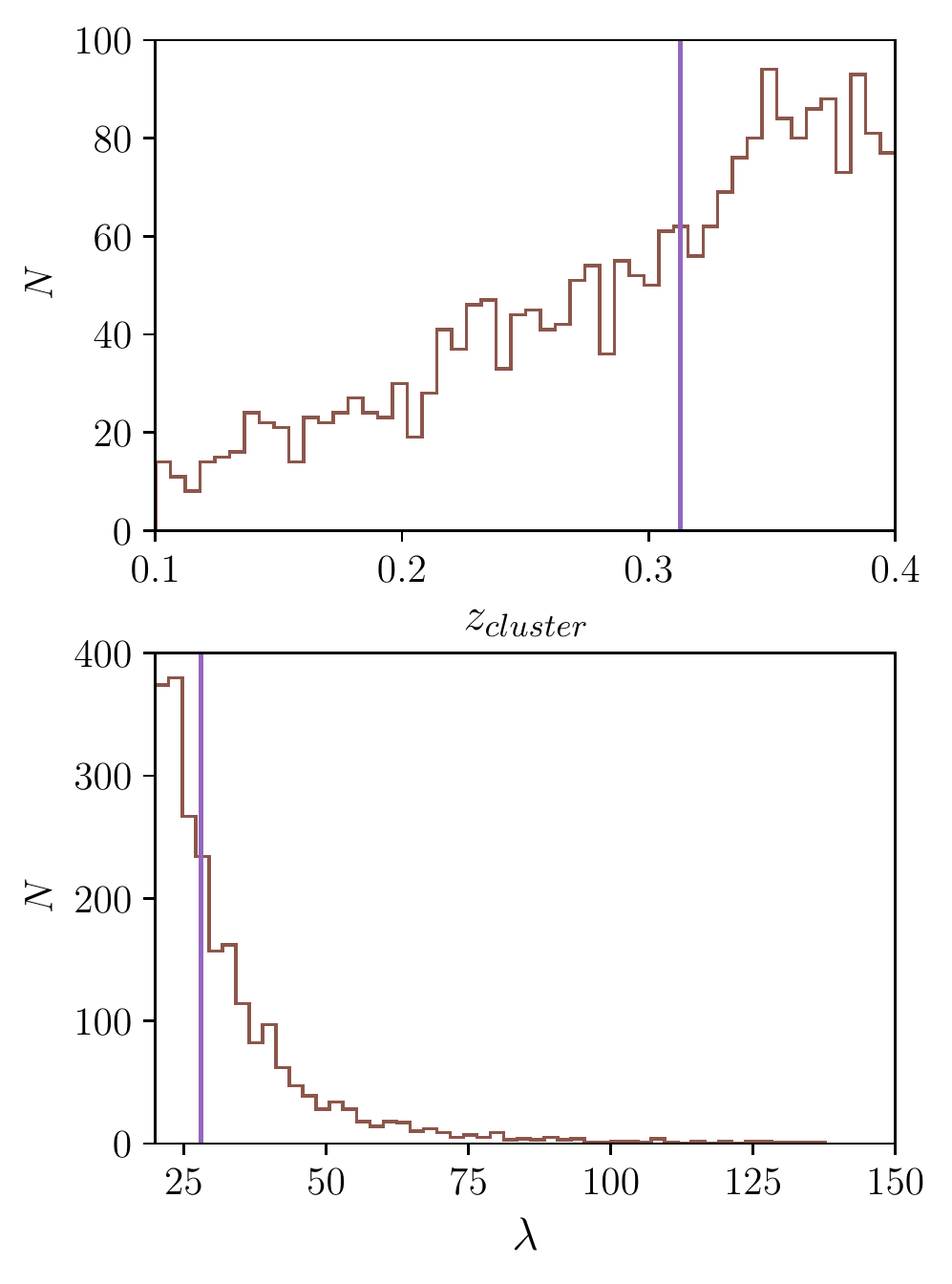}
    \caption{Distribution of redshifts (upper panel) and richness (lower panel) of
    the redMaPPer clusters used in this work. The vertical lines correspond to the median
    values of these parameters.}
    \label{fig:hist}
\end{figure}

The full sample consists on 26\,111 clusters with richness $20 \lesssim \lambda < 300$ spanning
a redshift range of $0.08 \lesssim z_c < 0.6$. For this work we restrict our sample to
$20 \leq \lambda < 150$ and $0.1 \leq z_c < 0.4$. 
The upper limit in the redshift 
selection is performed considering our weak-lensing data and because up to this redshift the number of galaxies lost due to the SDSS depth is relatively small. We also discard the clusters at the edges of the SDSS field, in order to avoid clusters with missing cluster members due to border issues. 
Therefore, we kept only the clusters that lie at more than $2$\,Mpc from the border of the
SDSS sky coverage. Finally, we only include the clusters located within the regions of the weak lensing data described previously, using the data from the deepest lensing catalog in the case of overlapping areas.
%
The total cluster sample used in this work includes 2\,275 clusters. 
We show in Fig. \ref{fig:hist} the distribution of richness and redshifts of these clusters.
To study the relation between the projected ellipticity and the mass and redshift of the halo, we split our total sample in four roughly equally number samples according to the median redshift ($z = 0.313$) and richness ($\lambda = 27.982$) of the sample.

\begin{figure*}
    \centering
    \includegraphics[scale=0.8]{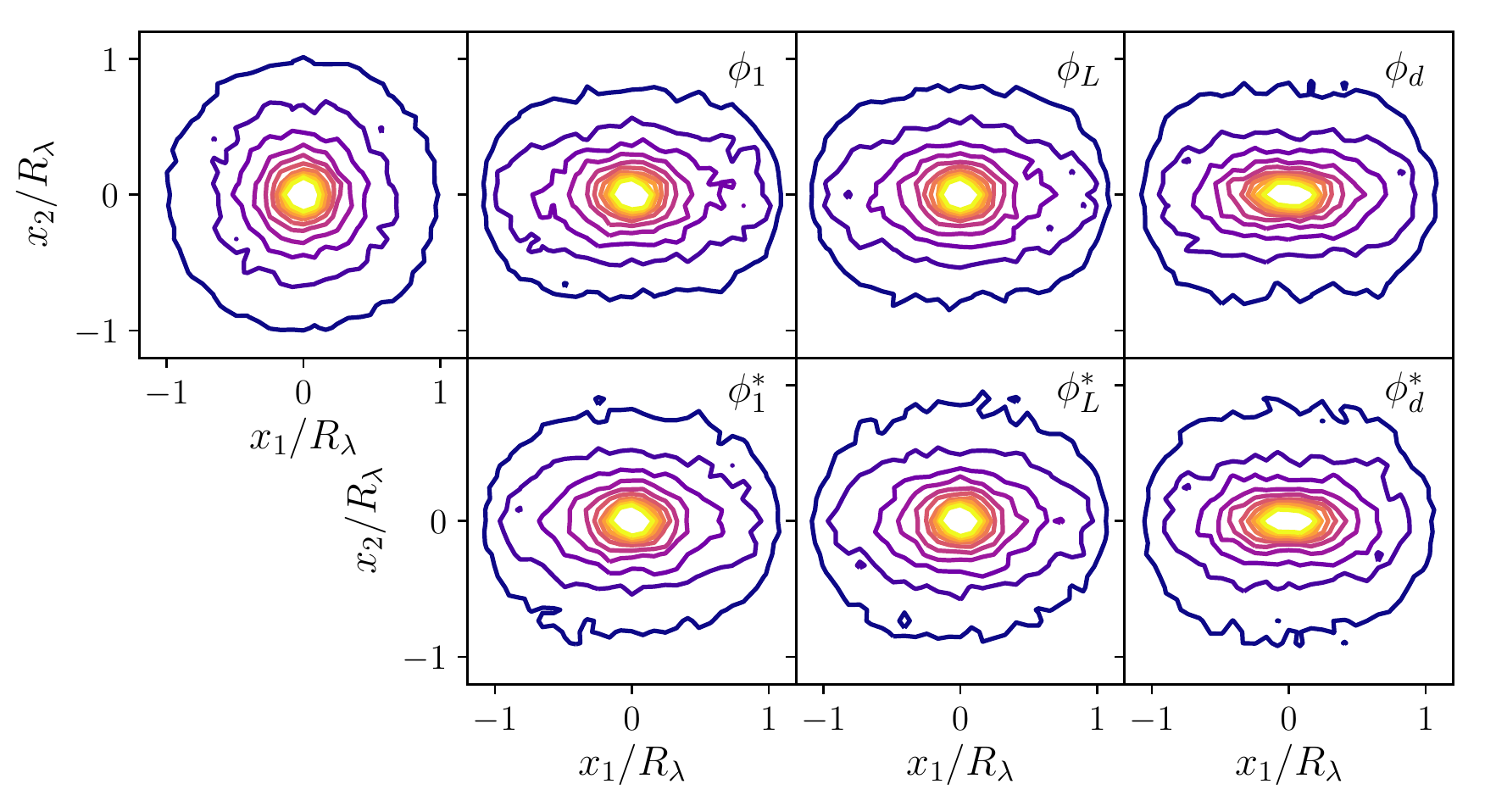}
    \caption{Satellite number density contours for the sample of redMaPPer clusters used in this work, obtained by stacking the galaxies aligned according to the estimated orientation angle. Projected positions, $x_1$ and $x_2$, are re-scaled with the cluster radius, $R_\lambda$.
    The first panel at the left was obtained without aligning the clusters. First row of panels
    is obtained considering all the satellites in the catalogue and the second row discarding the galaxies
    with $p_{mem} < 0.5$ to estimate the orientation angle. Second, third and fourth columns correspond 
    to the orientation computed considering a
    uniform, a luminosity and a distance weight to compute the quadrupole moments, respectively.}
    \label{fig:contours}
\end{figure*}

\section{Estimating the surface density orientation}
\label{sec:orientation}

In order to constrain the projected surface mass ellipticity,
it is necessary to know the orientation of the total mass distribution,
mostly traced by the dark matter, for each galaxy cluster. 
Unfortunately, these orientations are unknown and some proxies 
are needed in order to estimate them. In this work  
we assume that the galaxy cluster surface density distribution 
can be traced by the satellite distribution.
Following a similar approach as \citet{Shin2018} and \citet{Uitert2017},
we compute a raw ellipticity and a position
angle for each redMapper cluster according to the quadrupole moments
defined as:
\begin{equation}
    Q_{ij} = \frac{\sum_k (x_{i,k} x_{j,k} w_k)}{\sum_k w_k},
\end{equation}
where the sum runs over the $k$ satellite galaxies of the cluster,
$(x_{1,k},x_{2,k})$ are their projected coordinates in the image plane
with respect to the adopted cluster centre and $w_k$ is a weight 
assigned to each satellite.

For each cluster we can obtain the raw ellipticity
components as:
\begin{equation}
    \epsilon^{sat}_1 = \frac{Q_{11} -  Q_{22}}{Q_{11} +  Q_{22}},
\end{equation}
\begin{equation*}
    \epsilon^{sat}_2 = \frac{2 Q_{12}}{Q_{11} +  Q_{22}},
\end{equation*}
and the position angle of the major axis relative to the x-axis:
\begin{equation} \label{eq:phi}
    \tan{2 \phi} = \frac{2 Q_{12}}{Q_{11} -  Q_{22}}.
\end{equation}

To get an insight on which member galaxies
trace better the mass distribution,
we consider different criteria to compute the quadrupole moments. 
For the weights we consider three types: 
(1) a uniform weight, $w_k = 1$;
(2) according to the $r-$band luminosity, $w_k = L_k$ and
(3) according to the projected distance from the centre,
$w_k = 1/(x^2_{1,k} + x^2_{2,k})$.
We also take into account for the computation all the
galaxy members in the sample and only those with a membership
probability larger than $0.5$ ($p_{mem} > 0.5$). 
According to the different criteria adopted to compute the
quadrupole moments, we use in total 
six different proxies to estimate the orientation angle for each cluster, 
named as:
$\phi_1$, $\phi_L$, $\phi_d$, when considering the total sample
of satellites with an uniform weight, a luminosity 
weight and a distance weight, respectively, and $\phi^*_1$, $\phi^*_L$, $\phi^*_d$
when considering only the satellites with $p_{mem} > 0.5$ and the respective
weights. The membership probability cut applied to select the satellites, 
lowers the number of interlopers that could result in an
underestimated ellipticity measurement \citep{Shin2018},
i.e. lowers the number of foreground and background galaxies that were
wrongly classified as cluster members.
Nevertheless, since this parameter strongly depends on
the distance to the cluster centre, it can filter the satellites
that lie on the outskirt of the cluster. 

\begin{figure}
    \centering
    \includegraphics[scale=0.7]{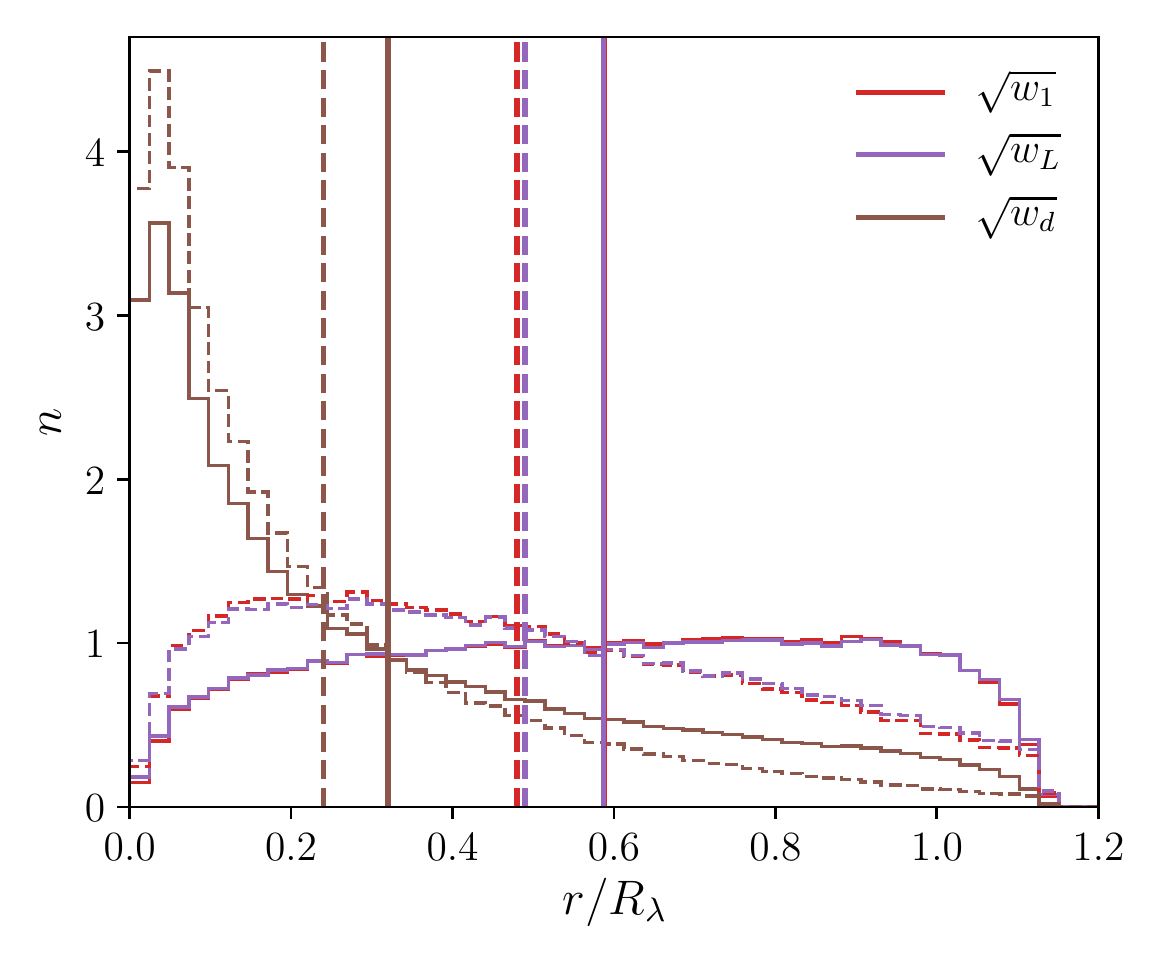}
    \caption{Normalized distributions of distances to the centre for the satellite galaxies, obtained
    by weighting the distances according to the square root of the defined uniform (red line), luminosity (purple line) and distance (brown line) weights. Dashed lines
    correspond to the distribution of satellites that satisfy $p_{mem} > 0.5$. Vertical lines show the 
    average weighted values for these distributions. Distributions for the total sample of satellites considering the uniform and luminosity weights are almost coincident, therefore the mean values are overlapped.}
    \label{fig:distances}
\end{figure}

In Fig. \ref{fig:contours} we show the number density contours
of all the satellites stacked along the different
orientations defined above, considering the projected coordinates
re-scaled according to the radius cluster, $R_\lambda$. In Fig. \ref{fig:distances}
we show the distribution of distances to the adopted cluster
centre for the satellites, derived weighting the distances according to the weights defined above, 
for the total sample of satellites and those with $p_{mem} > 0.5$.
As it can be noticed, when the distance weight 
is considered, the derived orientation angle follows the distribution of the
satellites that are located at the inner regions of the cluster. Also, as 
expected, when we consider only the satellites with $p_{mem} > 0.5$,
the distance distribution tend to lower values. In Table \ref{tab:ellipticities}
we show the estimated raw ellipticity values. Although with a high dispersion, 
larger values are derived
when considering a Luminosity weight and only satellites with $p_{mem} > 0.5$
to compute the quadrupole moments. Distribution of the differences 
between the orientation angles considering the defined weights and the membership
cut have a standard deviation of $\sim45-55$\,deg.

The satellite distribution has been used to estimate the halo ellipticity
in previous works \citep{Brainerd2005,Bailin2008,Shin2018}. As pointed out by \citet{Shin2018}, ellipticity measurements derived from the satellite distribution are affected by a noise bias, introduced by
the fact that a finite number of satellites is considered; an edge bias,
since members are selected within a circular aperture and a bias introduced by
the inclusion of interlopers, foreground or background red galaxies considered 
as members. In this work we do not intend to measure the halo ellipticity from
the satellite distribution but to use them as a proxy for the surface density orientation. 
Nevertheless, it is important to take into account that all the mentioned biases
can impact on the estimation of the orientation angle. These issues will be discussed in more detail in Sec. \ref{sec:bias}. 

\begin{table}
    \centering
    \begin{tabular}{c c c c c}
    \hline
    \hline
    Orientation & Satellite  & $w_k$   &  $\langle \epsilon^{sat} \rangle$ & $\sigma_\epsilon$ \\
                &   sample   &         & \\    
    \hline
$\phi_1$ & Total           &  $1$  & $0.21$ & $0.11$ \\
$\phi_L$ & Total           &   $L_k$    & $0.26$ & $0.13$ \\
$\phi_d$ & Total           &    $(x^2_{1,k} + x^2_{2,k})^{-1}$   & $0.18$ & $0.09$ \\
$\phi^*_1$ & $p_{mem} > 0.5$ &  $1$  & $0.29$ & $0.15$ \\
$\phi^*_L$ & $p_{mem} > 0.5$ &   $L_k$    & $0.35$ & $0.17$ \\
$\phi^*_d$ & $p_{mem} > 0.5$ &   $(x^2_{1,k} + x^2_{2,k})^{-1}$    & $0.22$ & $0.11$ \\
  \end{tabular}
    \caption{Mean raw ellipticity ($\langle \epsilon \rangle$) and standard deviation ($\sigma_\epsilon$)
    estimates for the sample of clusters, derived according to the quadrupole moments taking into account
    different samples of satellites and weights ($w_k$). }
    \label{tab:ellipticities}
\end{table}

\section{Weak lensing analysis}
\label{sec:analysis}

The weak gravitational lensing effect introduces a distortion in the luminous sources
that are behind a gravitational potential considered as the lens. 
Galaxy systems in particular are powerful lenses that distort
the shape of the galaxies that are behind, inducing an alignment of their shapes
along the direction tangential to the lens mass distribution \citep{Bartelmann2001}. 
This distortion can be quantified by the complex-value lensing shear,
$\gamma = \gamma_1 + i\gamma_2$, which can be estimated according to the measured ellipticity
of the background galaxies, i.e. the galaxies that lie behind the
galaxy cluster and thus, that are affected by the lensing effect. Since galaxies
have their own intrinsic ellipticity, the observed source shape
results in a combination of their intrinsic ellipticiy and the ellipticity
introduced by the lensing effect. Assuming that the galaxies are randomly orientated
in the sky (for a large sample of galaxies at a wide range of distances from the lens),
the shear can be estimated by averaging the ellipticity of many sources, $\langle e \rangle = {\gamma}$,
where the noise of this estimate scales with the number of sources ($N_S$)
considered in the average as $1/\sqrt{N_S}$.

In order to reduce the noise introduced by the intrinsic shape
of the considered sources we use stacking techniques
that consist on combining several lenses which artificially 
increase the density of sources. Stacking techniques can provide
a lensing signal with a suitable confidence level which allow to 
derive the average projected density distribution of the combined lenses 
\citep[e.g.][]{Foex2014, Leauthaud2017, Simet2017, Pereira2018,Chalela2018}. 
Usually, galaxy clusters are stacked without considering a particular orientation,
therefore, the derived radial density profiles
can be well fitted using axisymmetric density mass
distribution models. 

If the projected density distribution is elongated in a preferential direction,
this will cause an azimuthal variation in the lensing signal which can be modelled
in order to estimate the projected ellipticity. In that case, if the clusters are stacked 
taking into account the orientation of this preferential direction, 
the surface density, $\Sigma$, can be fitted
by considering a multipole expansion. A good aproximation is to model $\Sigma$
by decomposing it in two terms. The first term, the monopole, 
contains all the isotropic information regarding the mass distribution.
The second term, the quadrupole,
is proportional to the projected average aligned ellipticity component, $\epsilon$. 
Since the true orientation of the projected total mass distribution
is unknown, our estimate will be related to the true projected ellipticity component of
the surface density mass distribution, $\epsilon_{\rm{T}}$, through
(Schrabback et al. in prep):
\begin{equation} \label{eq:ellip}
    \epsilon = \epsilon_{\rm{T}} \cos{2 \Delta \phi},
\end{equation}
where $\Delta \phi$ is the misalignment angle between the true orientation and the proxies
defined in Sec. \ref{sec:orientation}. Therefore, our measured ellipticity will be 
underestimated and we expect higher $\epsilon$ values when the selected proxy is better aligned
with the mass distribution.

In this section, we first describe how the monopole is modelled
to derive the total masses of the considered cluster samples (subsection \ref{subsec:mono}).
Then, we fit the $\epsilon$ considering the quadrupole modelling described
in subsecion \ref{subsec:quadru}. To estimate the monopole and quadrupole components
we combine the shape measurements of the background galaxy sample according to the estimators defined in subsection \ref{subsec:estimators}

\subsection{Isotropic lens model}
\label{subsec:mono}

For any distribution of projected density mass, we can relate 
the azimuthally averaged tangential component of the shear, $\gamma_{\rm{t}}$,
with the contrast density distribution as \citep{Bartelmann1995}:
\begin{equation} \label{eq:DeltaSigma}
    \gamma_{\rm{t}}(r) \times \Sigma_{\rm crit} = \bar{\Sigma}(<r) -  \bar \Sigma(r)  \equiv \Delta \Sigma(r),
\end{equation}
where we define the density contrast, $\Delta\Sigma$. Here $\gamma_{\rm{t}}(r)$ is the tangential component of the shear at a projected distance from the centre of the mass distribution, $r$, $\bar{\Sigma}(<r) $ and $\bar \Sigma(r)$ are the azimuthally averaged projected surface density distribution within a disk and within a ring of distance $r$, respectively. $\Sigma_{\rm crit}$ is the critical density which contains all the geometrical information about the observer-lens-source configuration and is defined as:
\begin{equation} \label{eq:sig_crit}
\Sigma_{\rm{crit}} = \dfrac{c^{2}}{4 \pi G} \dfrac{D_{OS}}{D_{OL} D_{LS}},
\end{equation}
where $D_{OL}$, $D_{OS}$ and $D_{LS}$ are  the angular diameter distances from the observer to the lens, from the observer to the source and from the lens to the source, respectively.

On the other hand, the averaged cross-component of the shear, $\gamma_{\times}$, defined as the component tilted at $\pi$/4 relative to the tangential component, should be zero.  This quantity is therefore
commonly used as a null test to check for the presence of systematics in the data.

When averaging the tangential component within an annulus, the anisotropic components of the surface density are vanished and only the monopole
information is left (see subsection \ref{subsec:quadru}). To derive the total masses of the considered cluster sample, we model the monopole component taking into account two terms: a perfectly centred dark matter halo profile and a miscentring term. The second term considers the offset between the redMaPPer centre and their host dark matter halo. This modeling is motivated by the fact that we do not know the true location of the halo centre and an observationally-motivated centre is adopted. This results in an offset distribution that can be modeled with two populations, a mis-centred group and a well-centred group. According to hydrodynamic simulations, when galactic centres are adopted the well-centred group could include up to about $60\%$ of all the clusters \citep{Yan2020}. 

This miscentring affects the 
observed shear profile, flattening the lensing signal at the inner regions. Therefore, if this term is not considered, the inferred lensing masses can be underestimated by $\sim 30\%$ \citep{Yang2006,Johnston2007,Ford2014}.
The miscentring term is modelled following \citet{Yang2006,Johnston2007,Ford2014}. 
If we consider an axis-symmetric surface mass density distribution
whose centre is offset by $r_{s}$ with respect to the adopted centre in the lens plane, the observed
projected density profile will be:
\begin{equation}
    \Sigma(r|r_s) = \frac{1}{2\pi} \int^{2\pi}_0 \Sigma \left( \sqrt{r^2 + r^2_s + 2 r r_s \cos{\theta}}. \right) d\theta
\end{equation}
We adopt as the halo centre the centre provided by the redMaPPer algorithm. 
According to X-ray observations, two distinct cluster populations are distinguished in the 
offset distribution between the X-ray and redMaPPer galaxy centres.
The first population, considered as well-centred clusters,
includes $\sim80\%$ of the X-ray clusters 
where the X-ray centroid and the position of the central galaxy are 
in agreement within $\sim 50$kpc. The remaining $20\%$ of the galaxy clusters are merging systems
where the gas is significantly offset from the central galaxy \citep{Rozo2014}.
This fraction of clusters is expected to be shifted following a Gaussian distribution,
\begin{equation}
\label{eq:Pdist}
    P(r_s) = \frac{r_s}{\sigma^2_{off}} \exp{\left(-\frac{1}{2}\frac{r^2_{s}}{\sigma^2_{off}}\right)}.
\end{equation}
Taking this into account the miscentring term can be computed as follow:
\begin{equation} \label{eq:off}
   \bar \Sigma_{miss}(r) = \int_{0}^\infty P(r_s) \Sigma(r|r_s) dr,
\end{equation}
such that the miscentring term for the density contrast profile:
\begin{equation}
    \Delta \Sigma_{miss}(r) = \bar{\Sigma}_{miss}(<r) - \bar \Sigma_{miss}(r).
\end{equation}
The monopole component is therefore modelled as:
\begin{equation} \label{eq:monomodel}
    \Delta \Sigma = p_{cc}\Delta \Sigma_{cen} + (1-p_{cc}) \Delta \Sigma_{miss},
\end{equation}
where $p_{cc}$ is the fraction of well-centred clusters and $\Delta \Sigma_{cen}$ is obtained as:
\begin{equation}
  \Delta \Sigma_{cen}(r) = \bar{\Sigma}(<r) - \bar \Sigma(r).
\end{equation}

For the surface density distribution, $\Sigma$, we adopt a spherically symmetric NFW profile \citep{Navarro97}, which depends on two parameters, the radius that encloses a mean density equal to 200 times the critical density of the Universe, $R_{200}$, and a dimensionless concentration parameter, $c_{200}$. This density profile is given by:
\begin{equation} \label{eq:nfw}
\rho(r) =  \dfrac{\rho_{\rm crit} \delta_{c}}{(r/r_{s})(1+r/r_{s})^{2}},
\end{equation}
where  $r_{s}$ is the scale radius, $r_{s} = R_{200}/c_{200}$, $\rho_{\rm crit}$ is the critical density of the Universe at the mean redshift ($\langle z_c \rangle$) of the sample of stacked clusters and
$\delta_{c}$ is the cha\-rac\-te\-ris\-tic overdensity of the halo:
\begin{equation}
\delta_{c} = \frac{200}{3} \dfrac{c_{200}^{3}}{\ln(1+c_{200})-c_{200}/(1+c_{200})}.  
\end{equation}
We compute $\langle z_c \rangle$ by averaging the redshifts of the clusters in the sample weighted according to the number of source galaxies considered for each cluster to compute the profile.
The mass within $R_{200}$ can be obtained as \mbox{$M_{200}=200\,\rho_{\rm crit} (4/3) \pi\,R_{200}^{3}$}. In order to model the profile we use the lensing formulae presented by \citet{Wright2000}. 
In the fitting procedure we follow \citet{Uitert2012,Kettula2015} and \citet{Pereira2018}, by using a fixed mass-concentration relation $c_{200}(M_{200},z)$, derived from simulations by \citet{Duffy2008}: 
\begin{equation}
c_{200}=5.71\left(M_{200}/2 \times 10^{12} h^{-1}\right)^{-0.084}(1+\langle z_c \rangle)^{-0.47}.
\end{equation}

To estimate the masses
we fit the adopted model with two free parameters, $p_{cc}$ and $M_{200}$.
We fix the width of the offset distribution in
Eq. \ref{eq:Pdist}, $\sigma_{off} = 0.4 h^{-1}$Mpc, 
according to the result presented in \citet{Simet2017}.
We do not take into account a point mass term for a possible stellar-
mass contribution of the central galaxies and a so-called 2-halo
term due to neighbouring halos. In order to avoid their contributions
to the computed profiles, we fit the profiles from $100 h_{70}^{-1}$kpc
up to $5 h_{70}^{-1}$Mpc, were these terms are not expected to have a significant
impact for the mass estimates \citep{Mandelbaum2006,Simet2017,Pereira2018}.

\subsection{Anisotropic lens model}
\label{subsec:quadru}

We follow \citet{Uitert2017} and model the observed lensing signal of the clusters,
by a mass distribution with confocal elliptical isodensity
contours of axis-ratio $q \leq 1$.  The surface density
of elliptical halos can be modelled considering a multipole expansion \citep{Schneider1991}
of $\Sigma(R)$ in terms of the ellipticity defined as $\epsilon:= (1-q)/(1+q)$:
\begin{equation}
\label{eq:Smyq}
\Sigma(R) = \Sigma(r,\theta) := \Sigma_0(r) + \epsilon \Sigma_2(r) \cos(2\theta).
\end{equation}
Here we neglected the higher order terms in $\epsilon$. This parameter is the estimated average aligned ellipticity component which is related to the true ellipticity of the projected total mass distribution through Eq. \ref{eq:ellip}. $R$ is the ellipsodal 
radial coordinate, $R^2 = r^2(q \cos^2(\theta) + \sin^2(\theta)/q)$
and $\theta$ is the position angle of the source relative to the major
semi-axis of the surface density distribution. $\Sigma_0$ is the monopole for which we consider an NFW distribution. The quadrupole is defined in terms of the monopole as $\Sigma_2 = -r d(\Sigma_0(r))/dr$.

The tangential and cross shear components can be also decomposed into the monopole
and quadrupole contributions:
\begin{align} \label{eq:gamma}
&    \gamma_{\rm{t}} (r,\theta) = \gamma_{\rm{t},0}(r) + \epsilon \gamma_{\rm{t},2}(r) \cos(2\theta),\\
&    \gamma_\times (r,\theta) = \epsilon \gamma_{\times,2}(r) \sin(2\theta). \nonumber
\end{align}
The tangential components of the monopole, $ \gamma_{\rm{t},0}$,
and the quadrupole, $\gamma_{\rm{t},2}$, as well as the cross component of the
quadrupole $\gamma_{\times,2}$ can be obtained as \citep{Uitert2017}:
\begin{align} \label{eq:gcomponents}
& \Sigma_{crit}   \gamma_{\rm{t},0}(r)  = \frac{2}{r^2} \int^r_0 r^\prime \Sigma_0(r^\prime) dr^\prime - \Sigma_0(r),\\
& \Sigma_{crit} \gamma_{\rm{t},2}(r) = -\frac{6 \psi_2(r)}{r^2} - 2\Sigma_0(r) - \Sigma_2(r) \nonumber, \\
& \Sigma_{crit} \gamma_{\times,2}(r) = -\frac{6 \psi_2(r)}{r^2} - 4\Sigma_0(r),  \nonumber
\end{align}
where $\psi_2(r)$ is the quadrupole component of the lensing potential and is obtained as:
\begin{equation}\label{eq:psi2}
    \psi_2(r) = -\frac{2}{r^2} \int_0^r r^{\prime 3} \Sigma_0(r^\prime) dr^{\prime}.
\end{equation}
We multiply the shear components by the critical density defined
in Eq. \ref{eq:sig_crit} so that we can combine the observed signal
at different source and lens redshifts, obtaining distance independent quantities.
The tangential component of the monopole is the usual contrast density
 $\Delta \Sigma$ defined in Eq. \ref{eq:DeltaSigma},
\begin{equation}
\Delta \Sigma(r) =  \Sigma_{crit} \gamma_{\rm{t},0}(r) = \frac{1}{2\pi} \int_0^{2\pi} \Sigma_{crit}, \gamma_{\rm{t}} (r,\theta) d\theta   
\end{equation}
which is the only term observed in the case of an axisymmetric mass distribution.

If we average the projected tangential and cross components in annular bins,
we obtain only the quadrupole components scaled according to the ellipticity:
\begin{align} \label{eq:gproj}
& \Gamma_{\rm{t} \cos{2\theta}}(r) := \epsilon \Sigma_{crit} \gamma_{\rm{t},2}(r) = \frac{1}{\pi} \int_0^{2\pi} \Sigma_{crit} \gamma_{\rm{t}} (r,\theta) \cos(2\theta) d\theta, \\
& \Gamma_{\times \sin{2\theta}}(r) := \epsilon \Sigma_{crit} \gamma_{\times,2}(r) = \frac{1}{\pi} \int_0^{2\pi} \Sigma_{crit} \gamma_\times (r,\theta) \sin(2\theta) d\theta,
\end{align}
where we defined the distance independent quantities $\Gamma_{\rm{t} \cos{\theta}}$
and $\Gamma_{\times \sin{\theta}}$.


\subsection{Estimators and fitting procedure}
\label{subsec:estimators}

In order to obtain the monopole and quadrupole profiles, we use
a stacking procedure by combining the shape measurements of all 
the background galaxies selected for each sample of clusters, artificially increasing
the number density of background galaxies and lowering the noise
introduced by their intrinsic shapes.

We define the estimators as:
\begin{align} \label{eq:gproj}
& \Delta \tilde{\Sigma}(r) = \frac{\sum_{j=1}^{N_L} \sum_{i=1}^{N_{S,j}} \omega_{LS,ij} \Sigma_{{\rm crit},ij} e_{{\rm t},ij}}{\sum_{j=1}^{N_L} \sum_{i=1}^{N_{S,j}} \omega_{LS,ij}}, \\
& \tilde{\Gamma}_{\rm{t} \cos{2\theta}}(r) = \frac{\sum_{j=1}^{N_L} \sum_{i=1}^{N_{S,j}} \omega_{LS,ij}  \Sigma_{{\rm crit},ij} e_{{\rm t},ij} \cos{2\theta}}{\sum_{j=1}^{N_L} \sum_{i=1}^{N_{S,j}} \omega_{LS,ij} \cos^2{2\theta}}, \\
& \tilde{\Gamma}_{\times \sin{2\theta}}(r) = \frac{\sum_{j=1}^{N_L} \sum_{i=1}^{N_{S,j}} \omega_{LS,ij}  \Sigma_{{\rm crit},ij} e_{{\rm t},ij} \sin{2\theta}}{\sum_{j=1}^{N_L} \sum_{i=1}^{N_{S,j}} \omega_{LS,ij} \sin^2{2\theta}},
\end{align}
where $\omega_{LS,ij}$ is the inverse variance weight computed according to the weight, $\omega_{ij}$, given by the $lens$fit algorithm for each background galaxy, $\omega_{LS,ij}=\omega_{ij}/\Sigma^2_{{\rm crit},ij}$. $N_L$ is the number of clusters considered for the stacking and $N_{S,j}$ the number of background galaxies located at a distance $r \pm \delta r$ from the $j$th cluster. $\Sigma_{{\rm crit},ij}$ is the critical density for the $i-$th source of the $j-$th lens.

We take into account a noise bias factor correction as suggested by \citet{Miller2013}, which considers the multiplicative shear calibration factor $m(\nu_{SN},l)$ provided by \textit{lens}fit. For this correction we compute:
\begin{equation}
1+K(z_L)= \frac{\sum_{j=1}^{N_L} \sum_{i=1}^{N_{S,j}} \omega_{LS,ij} (1+m(\nu_{SN,ij},l_{ij}))}{\sum_{j=1}^{N_{Lens}} \sum_{i=1}^{N_{S,j}} \omega_{LS,ij}},
\end{equation}
following \citet{Velander2014,Hudson2015,Shan2017,Leauthaud2017,Pereira2018} we calibrate our estimators multiplying them by a 
factor $(1+K(z_L))^{-1}$. The expectation values of the defined calibrated estimators will be:
\begin{align} \label{eq:gproj}
& \rm{E} \left( \frac{\Delta \tilde{\Sigma}(r)}{1+K(z_L)} \right)  = \Delta \Sigma(r), \\
& \rm{E} \left( \frac{\tilde{\Gamma}_{\rm{t} \cos{2\theta}}(r)}{1+K(z_L)} \right)  = \Gamma_{\rm{t} \cos{2\theta}}(r), \\
& \rm{E} \left( \frac{\tilde{\Gamma}_{\times \sin{2\theta}}(r)}{1+K(z_L)} \right)  = \Gamma_{\times \sin{2\theta}}(r).
\end{align}
Using these estimators we compute the profiles using 10 non-overlapping concentric logarithmic annuli to preserve the signal-to-noise ratio of the outer region, from $100\,h^{-1}_{70}$\,kpc up to $5.0\,h^{-1}_{70}$\,Mpc. Given that the uncertainties in the estimated lensing signal are expected to be dominated by shape noise, we do not expect a noticeable covariance between adjacent radial bins and so we treat them as independent in our analysis. Accordingly, we compute error bars for each radial bin of the monopole and quadrupole profiles by bootstrapping the lensing signal using 100 realizations. 

The parameters that are fitted from the defined estimators are the logarithmic mass, $\log(M_{200})$, the fraction of miscentring clusters, $p_{cc}$ and the average aligned ellipticity component, $\epsilon$. Our procedure consists of fitting $\log(M_{200})$ and $p_{cc}$, considering only the monopole component ($\Delta \Sigma(r)$, see Eq. \ref{eq:monomodel}) according to the modelling presented in subsection \ref{subsec:mono}. Then, taking into account the estimated $M_{200}$ we constrain $\epsilon$ by simultaneously fitting the quadrupole components, $\Gamma_{\rm{t} \cos{2\theta}}(r)$ and $\Gamma_{\times \sin{2\theta}}(r)$. To model the anisotropic contribution of the shear-field included in the quadrupole components, we neglect the miscentring term, in spite it is considered to derive the halo masses by fitting the density contrast profile (Eq. \ref{eq:monomodel}). A wrong determination of the halo centre, as it is expected for $\sim 20 \%$ of the redMapper clusters, can lead to underestimated lensing masses up to $30\%$, which would result in an overestimation of the projected ellipticities in $\sim 25 \%$. Nevertheless, we do not expect a significant
contribution of this term in the quadrupole component as previously reported by other studies \citep{Uitert2017,Shin2018}. 

We constrain our free parameters by using the Markov chain Monte Carlo (MCMC) method, implemented through \texttt{emcee} python package \citep{Foreman2013}, to optimize the  log-likelihood functions for the monopole profile, $\ln{\mathcal{L}}(\Delta \Sigma | r ,M_{200},p_{cc})$ and for the quadrupoles, $\ln{\mathcal{L}}(\Gamma_{\rm{t} \cos{2\theta}} | r ,\epsilon) +  \ln{\mathcal{L}}(\Gamma_{\times \sin{2\theta}} | r ,\epsilon)$. 
To fit the data we use 10 chains for each parameter and 200 steps, considering flat priors for $\log(M_{200})$ and $\epsilon$, ($12.5 < \log(M_{200}/(h^{-1}_{70} M_\odot)) < 15.5$  and $0.0 < \epsilon < 0.5$) and for $p_{cc}$ we take into account the results from the X-ray observations \citep{Rozo2014} and consider as a prior a Gaussian distribution with a mean value of 0.8 and a standard deviation of 0.3. Our best fit parameters are obtained according to the median of the posterior distributions and errors are based on the differences between the median and the $16^{th}$ and $84^{th}$ percentiles, after discarding the first 50 steps of each chain.

In order to obtain information of the projected ellipticities at different radial distances, quadrupole profiles are fitted within three ranges: $0.1 < r/(h^{-1}_{70}$\,Mpc)$ < 5$,
$0.1 < r/(h^{-1}_{70}$\,Mpc)$ < 0.7$ and $0.7 < r/(h^{-1}_{70}$\,Mpc)$ < 5$.
We refer to these ranges as $r_{0.1}^{5.0}$, $r_{0.1}^{0.7}$ and $r_{0.7}^{5.0}$, respectively.
The selection cut at $700\,h^{-1}_{70}$\,kpc was made to have
the same number of estimates in the inner and outer regions (5 points each). 
We expect that the constrained $\epsilon$ within $r_{0.1}^{0.7}$ to be more related to the projected density distribution at the inner regions of the clusters, thus, with the dark matter halo elongation. On the other hand, the fitted ellipticity within $r_{0.7}^{5.0}$ would be related to the mass distribution at the outskirts of the galaxy clusters.

\begin{figure}
    \centering
    \includegraphics[scale=0.8]{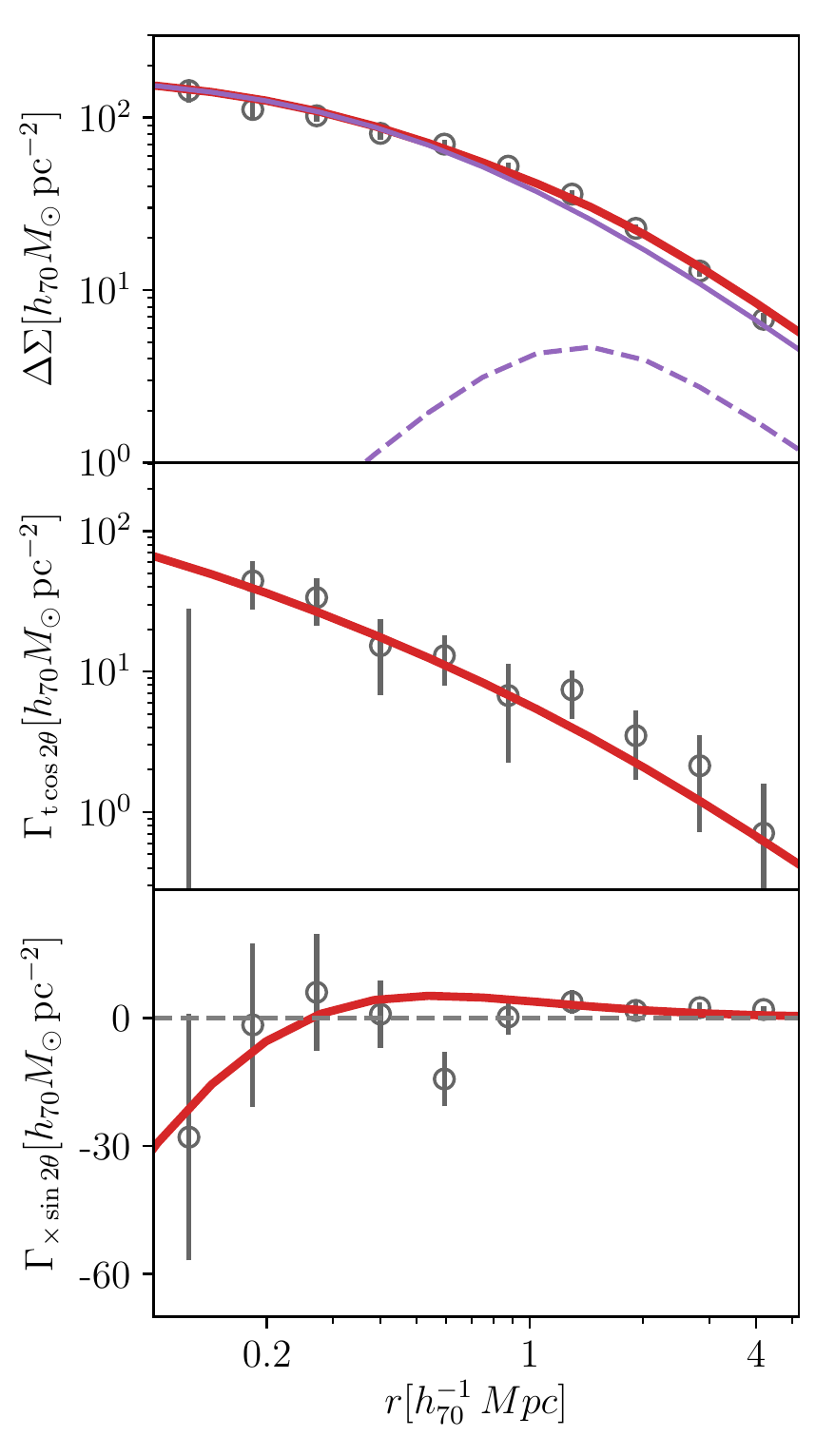}
    \caption{Monopole (upper panel) and quadrupole tangential (middle panel) and cross (bottom panel) components, computed for the total sample of clusters analysed in this work. Solid thick lines are the derived models according to the fitted mass, $M_{200}$, and  fraction of miscentring clusters, $p_{cc}$, for the monopole, and the average aligned ellipticity component, $\epsilon$, for the quadrupole components. In the upper panel we also show both fitted terms for the monopole, the centred $\Delta \Sigma_{cen}$ (solid purple line), and the miscentring term, $\Delta \Sigma_{miss}$ (dashed purple line), terms (see Eq. \ref{eq:monomodel}). }
    \label{fig:profile}
\end{figure}

We show as an example in Fig. \ref{fig:profile}, the computed monopole and quadrupole components for the total sample of redMaPPer, fitted within the whole projected distance range ($r_{0.1}^{5.0}$). The quadrupoles were obtained by considering $\phi_1$ as the proxy to estimate the orientation angle for the surface density distribution. The remaining  fitted quadrupole components derived for all the estimated orientations and all the considered cluster samples are shown in Appendix \ref{app:quadprofiles}. 

\section{Results}
\label{sec:results}

\begin{figure*}
    \centering
    \includegraphics[scale=0.65]{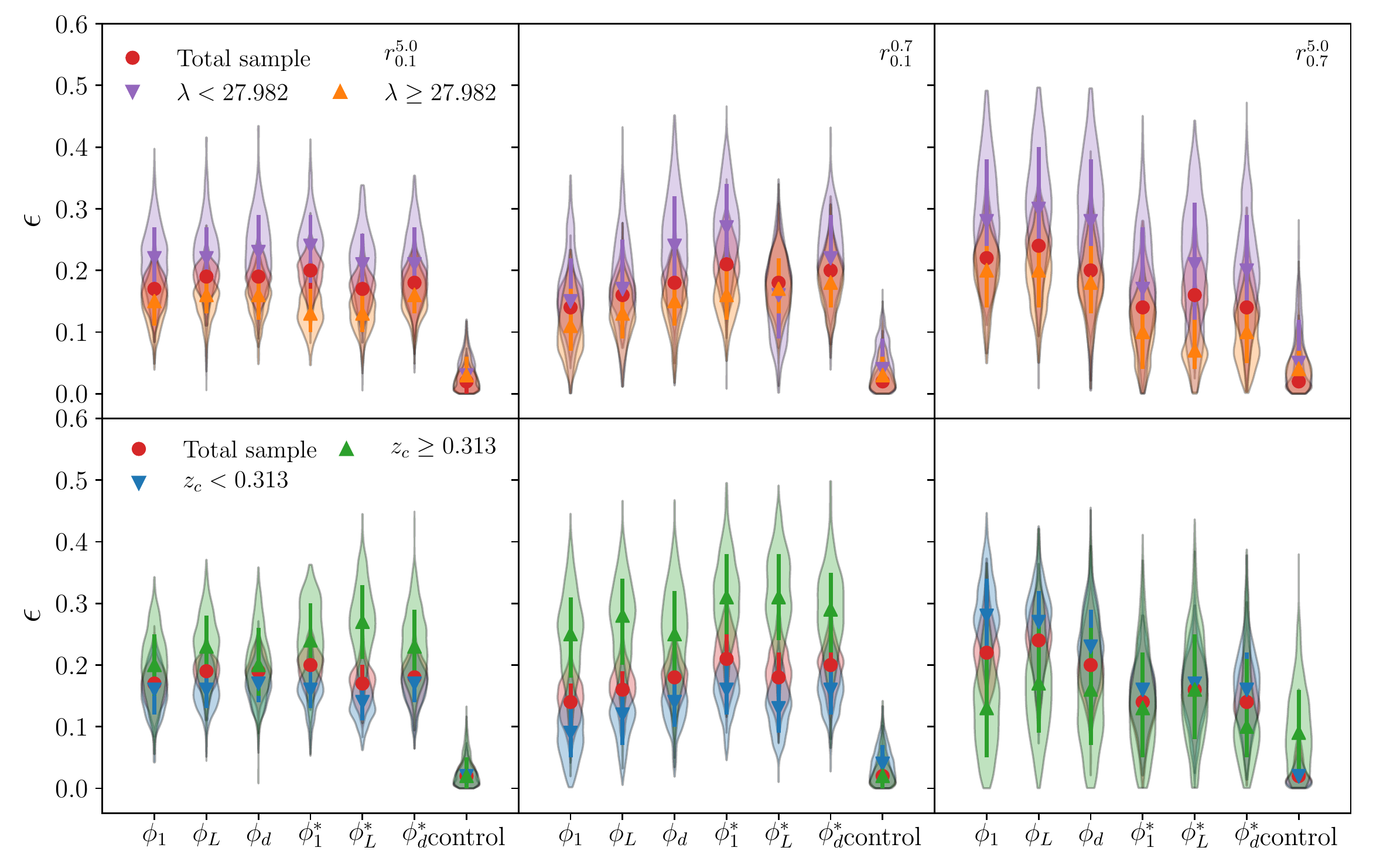}
    \caption{Posterior distributions of the average aligned ellipticity components, according to the different orientation angle criteria, for the total sample and low- and high-richness sample (upper panel) and for for the total sample and low- and high-redshift samples (lower panel). Symbols and bars correspond to the median and error bars to the $16^{th}$ and $84^{th}$ percentiles of the posterior distributions. Left, middle and right panels show the derived estimates fitting the quadrupole profiles within $0.1\,h^{-1}_{70}$\,Mpc$ < r < 5\,h^{-1}_{70}$\,Mpc, $0.1\,h^{-1}_{70}$\,Mpc$ < r < 0.7\,h^{-1}_{70}$\,Mpc and $0.7\,h^{-1}_{70}$\,Mpc$ < r < 5\,h^{-1}_{70}$\,Mpc, respectively.}
    \label{fig:emedians}
\end{figure*}

\begin{figure*}
    \centering
    \includegraphics[scale=0.65]{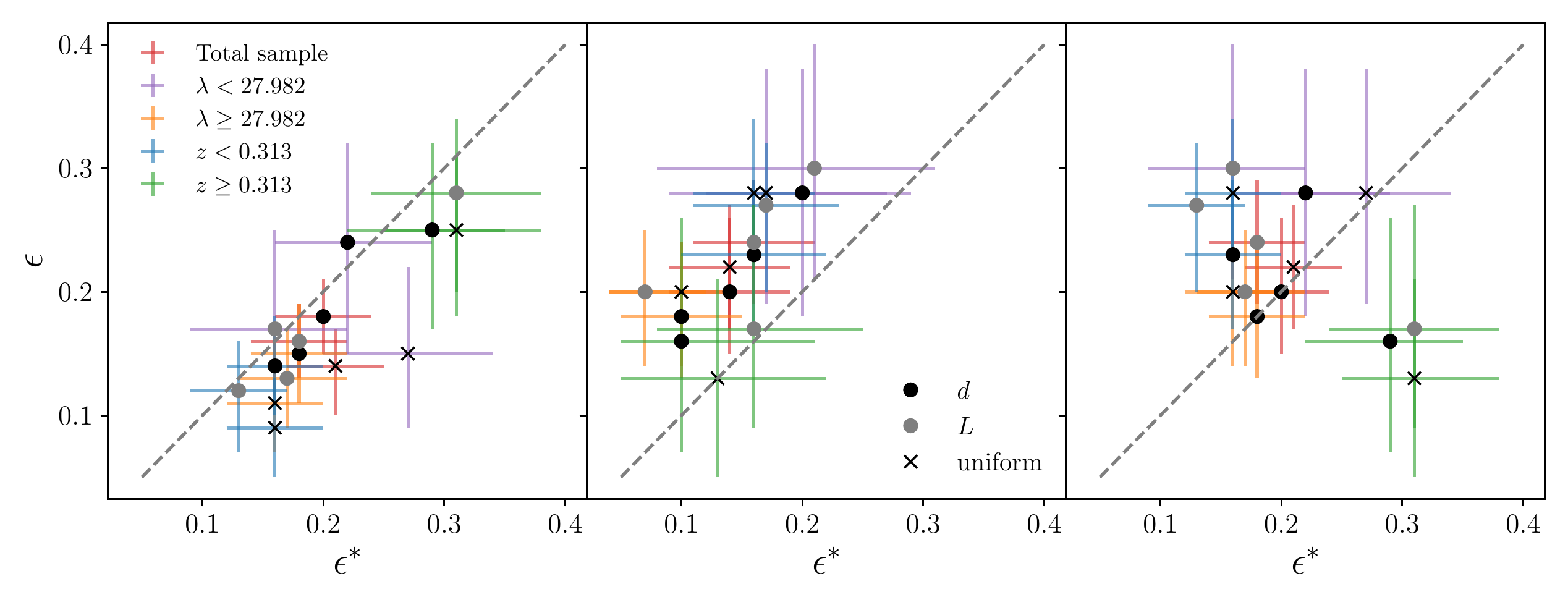}
    \caption{Derived average aligned ellipticity component estimates by fitting quadrupole profiles in the inner region ($r_{0.1}^{0.7}$, left panel) and in the outer region ($r_{0.7}^{5.0}$, middle panel). The right panel shows the ellipticities computed by fitting the quadrupoles in $r_{0.1}^{0.7}$ (x-axis) and  $r_{0.7}^{5.0}$ (y-axis) ranges. $\epsilon^*$ and $\epsilon$ refer to the ellipticity computed aligning the clusters according the satellites that satisfy the membership cut and the whole sample of satellites, respectively. Black and grey dots and crosses correspond to the quadrupoles computed considering the orientation angle weighting the satellites taking into account a distance, a luminosity and a uniform weight, as defined in Sec. \ref{sec:orientation}. Dashed gray line corresponds to the identity.}
    \label{fig:erranges}
\end{figure*}

We obtained the average aligned ellipticity components for the five samples of redMaPPer clusters (total , high- and low-redshift, high- and low-richness samples) by fitting both quadrupole component profiles in the mentioned projected distance ranges. In order to obtain the quadrupole profiles, we compute $\theta$ for each source considering the halo orientations derived for each cluster, $\phi$ (Eq. \ref{eq:phi}), taking into account the proxies defined in Sec. \ref{sec:orientation} ($\phi_1$,$\phi_L$,$\phi_d$,$\phi^*_1$,$\phi^*_L$ and $\phi^*_d$). We also compute a control quadrupole profile, considering that the halo is aligned with the R.A. axis, i.e., $\theta$ is taken as the position angle of the source with respect to the R.A. Since, we do not expect the halo orientation to be correlated with the sky-position, estimated ellipticities should be consistent with zero, as shown for the satellite distribution in the first panel of Fig. \ref{fig:contours}.  Resultant fitted parameters for the monopole and quadrupole terms are shown in Table \ref{table:results}. All the fitted ellipticities for the quadrupole control profiles are consistent with zero within $1.5\sigma$. 
In Fig. \ref{fig:emedians} we show the posterior distributions of the estimated ellipticities together with their medians and errors.  We can notice that the derived ellipticities are  higher when considering the sample of clusters with a lower mean mass according to the cut in richness ($\lambda < 27.982$), regardless of the proxy used to define the halo orientation. The average ellipticity for the higher mass sample ($\lambda \geq 27.982$) is $0.15 \pm 0.03$, while for the lower mass sample we obtain $0.22^{+0.06}_{-0.05}$ when considering the full range to fit the profile ($r_{0.1}^{5.0}$).  
This result does not follow the trend expected from $\Lambda$CDM simulations where higher mass halos tend to be less spherical. Nevertheless, our determinations are in agreement within $1\sigma$. We discuss this issue further in Sec. \ref{sec:discuss}. We also obtain larger projected ellipticity values for higher redshift clusters, in agreement with the expectations from $\Lambda$CDM numerical simulations.  Differences in the ellipticity estimates between high- and low-redshift clusters are larger when the orientation angle is obtained only taking into account satellites with a higher membership probability ($p_{mem}$ > 0.5). When the profiles are fitted in the whole projected distance range ($r_{0.1}^{5.0}$) and only the satellites with $p_{mem} > 0.5$ are included to derive the orientation angle, the derived average ellipticities are $0.16 \pm 0.03$ and $0.25^{+0.05}_{-0.06}$ for the low- and high-redshift sample of clusters, respectively. Differences are even larger when the profiles are fitted up to $0.7\,h^{-1}_{70}$Mpc, obtaining average ellipticities of $0.15 \pm 0.04$ and $0.30 \pm 0.07$, respectively. On the other hand, if the profiles are fitted only in the outer region, from $0.7\,h^{-1}_{70}$Mpc, differences are only significant when considering the full sample of satellites to trace the mass distribution and with low-redshift clusters showing the higher ellipticity values. For our high-redshift sample the fraction of satellites that satisfy $p_{mem}$ < 0.5 are $\sim 50\%$, while for the low-redshift is the $\sim 64\%$. Although the fraction of satellites with $p_{mem}$ < 0.5 is lower for the low-redshift sample, they contribute by tracing the mass distribution at the cluster outskirts. 

In Fig. \ref{fig:erranges} we compare the ellipticities obtained by fitting the quadrupole profiles in the inner and outer regions, and with and without considering the membership cut in the satellite sample to estimate the orientation angle. The general tendencies discussed for the high and low richness and redshift samples can be also noticed in this Figure. We do not observe a general tendency between the projected ellipticity estimates when the different weights (a uniform, a luminosity and a distance weight) are considered to derive the orientation angle of the mass distribution. Taking into account the large dispersion observed in the distribution of differences of the derived angles ($\sim 50$\,deg), the observed lack of impact of the weights on the ellipticity estimates, points to a general poorly orientation estimate of the mass distribution. Nevertheless, it can be noticed that the membership cut applied to the satellite samples used to compute the halo orientation, can affect differently the projected ellipticities in the inner and the outer regions of the quadrupole profile. To obtain information on how the membership cut in the computation of the surface density orientation affects the projected ellipticity estimates, we compute the ratio $\epsilon/\epsilon^*$, defined as the ratio between the projected ellipticity estimated considering the whole sample of satellites to derive the orientation angle, $\epsilon$, and using only the satellites that satisfy $p_{mem}$ > 0.5, $\epsilon^*$. For all the cluster samples, the proxy selection to define the halo orientation angle has a low impact on the derived projected ellipticities when the profiles are fitted over the whole projected radius range ($r_{0.1}^{5.0}$), being the average $\langle \epsilon/\epsilon^* \rangle = 1.01$ with a standard deviation of $0.12$. Nevertheless, when the profiles are fitted up to $700\,h^{-1}_{70}$kpc ($r_{0.1}^{0.7}$), projected ellipticities tend to be higher if only the satellites with $p_{mem} > 0.5$ are considered to estimate the orientation angle ($\langle \epsilon/\epsilon^* \rangle = 0.82$ with a standard deviation of $0.15$). This general tendency is reversed when the profile is fitted in the outer region ($\langle \epsilon/\epsilon^* \rangle = 1.60$ with a standard deviation of $0.41$). Therefore, the inner part of the halo is better traced by the satellites that have a higher probability of being members. On the other hand, the outer part of the  projected density distribution is better traced by the whole sample of satellites. 

\begin{table*}
\caption{Fitted parameters from the monopole and quadrupole profile components for the redMaPPer cluster samples.}
\begin{tabular}{c c | c c c | c | c c | c c | c c}
\hline
\hline
\rule{0pt}{1.05em}%
Clusters                       & $N_L $   &  $M_{200}$ & $p_{cc}$ & $\chi^2_{red}$  &  Orientation & \multicolumn{2}{c|}{$r_{0.1}^{5.0}$} & \multicolumn{2}{c|}{$r_{0.1}^{0.7}$} & \multicolumn{2}{c|}{$r_{0.7}^{5.0}$} \\
                               &                 &  [$h^{-1}_{70} 10^{14} M_\odot$]          &          &                 &  angle       & $\epsilon$ & $\chi^2_{red}$ &   $\epsilon$ & $\chi^2_{red}$ &   $\epsilon$ & $\chi^2_{red}$ \\
\hline
\rule{0pt}{1.05em}   
\vspace{0.15cm}
Total sample                   & 2275 &  $2.06_{-0.11}^{+0.09}$ & $0.79_{-0.04}^{+0.04}$ & 1.49 & $\phi_1$    & $0.17_{-0.03}^{+0.03}$ & 1.16 & $0.14_{-0.04}^{+0.03}$  & 1.22  & $0.22_{-0.05}^{+0.05}$ & 1.34 \\ \vspace{0.15cm}
                               &      &                         &                        &      & $\phi_L$    & $0.19_{-0.03}^{+0.02}$ & 1.90 & $0.16_{-0.03}^{+0.03}$  & 1.92  & $0.24_{-0.05}^{+0.05}$ & 2.10 \\ \vspace{0.15cm}
                               &      &                         &                        &      & $\phi_w$    & $0.19_{-0.03}^{+0.03}$ & 0.64 & $0.18_{-0.03}^{+0.03}$  & 0.65  & $0.20_{-0.05}^{+0.06}$ & 0.65 \\ \vspace{0.15cm}
                               &      &                         &                        &      & $\phi^*_1$  & $0.20_{-0.03}^{+0.03}$ & 1.04 & $0.21_{-0.04}^{+0.04}$  & 1.07  & $0.14_{-0.05}^{+0.05}$ & 1.17 \\ \vspace{0.15cm}
                               &      &                         &                        &      & $\phi^*_L$  & $0.17_{-0.03}^{+0.03}$ & 1.11 & $0.18_{-0.04}^{+0.04}$  & 1.11  & $0.16_{-0.05}^{+0.05}$ & 1.12 \\ \vspace{0.15cm}
                               &      &                         &                        &      & $\phi^*_w$  & $0.18_{-0.03}^{+0.03}$ & 0.78 & $0.20_{-0.04}^{+0.04}$  & 0.81  & $0.14_{-0.05}^{+0.05}$ & 0.87 \\ \vspace{0.15cm}
                               &      &                         &                        &      & control     & $0.02_{-0.02}^{+0.02}$ & 1.27 & $0.02_{-0.01}^{+0.02}$  & 1.26  & $0.02_{-0.01}^{+0.04}$ & 1.27 \\ \vspace{0.15cm}
$\lambda < 27.982$             & 1139 &  $1.43_{-0.13}^{+0.11}$ & $0.82_{-0.06}^{+0.06}$ & 0.91 & $\phi_1$    & $0.22_{-0.05}^{+0.05}$ & 1.16 & $0.15_{-0.06}^{+0.07}$  & 1.21  & $0.28_{-0.09}^{+0.10}$ & 1.25 \\ \vspace{0.15cm}
                               &      &                         &                        &      & $\phi_L$    & $0.22_{-0.05}^{+0.05}$ & 2.10 & $0.17_{-0.06}^{+0.08}$  & 2.15  & $0.30_{-0.09}^{+0.10}$ & 2.20 \\ \vspace{0.15cm}
                               &      &                         &                        &      & $\phi_w$    & $0.23_{-0.06}^{+0.06}$ & 0.93 & $0.24_{-0.09}^{+0.08}$  & 0.92  & $0.28_{-0.10}^{+0.10}$ & 0.96 \\ \vspace{0.15cm}
                               &      &                         &                        &      & $\phi^*_1$  & $0.24_{-0.06}^{+0.05}$ & 1.21 & $0.27_{-0.07}^{+0.07}$  & 1.22  & $0.17_{-0.08}^{+0.10}$ & 1.29 \\ \vspace{0.15cm}
                               &      &                         &                        &      & $\phi^*_L$  & $0.21_{-0.06}^{+0.05}$ & 1.35 & $0.16_{-0.07}^{+0.06}$  & 1.38  & $0.21_{-0.13}^{+0.10}$ & 1.35 \\ \vspace{0.15cm}
                               &      &                         &                        &      & $\phi^*_w$  & $0.21_{-0.05}^{+0.06}$ & 0.65 & $0.22_{-0.06}^{+0.07}$  & 0.65  & $0.20_{-0.08}^{+0.09}$ & 0.66 \\ \vspace{0.15cm}
                               &      &                         &                        &      & control     & $0.03_{-0.02}^{+0.03}$ & 1.03 & $0.04_{-0.03}^{+0.05}$  & 1.06  & $0.05_{-0.03}^{+0.07}$ & 1.07 \\ \vspace{0.15cm}
$\lambda \geq     27.982$      & 1136 &  $2.71_{-0.13}^{+0.13}$ & $0.78_{-0.04}^{+0.05}$ & 1.16 & $\phi_1$    & $0.15_{-0.04}^{+0.03}$ & 1.19 & $0.11_{-0.04}^{+0.06}$  & 1.25  & $0.20_{-0.06}^{+0.04}$ & 1.33 \\ \vspace{0.15cm}
                               &      &                         &                        &      & $\phi_L$    & $0.16_{-0.03}^{+0.03}$ & 1.37 & $0.13_{-0.04}^{+0.04}$  & 1.39  & $0.20_{-0.06}^{+0.05}$ & 1.46 \\ \vspace{0.15cm}
                               &      &                         &                        &      & $\phi_w$    & $0.16_{-0.04}^{+0.03}$ & 0.94 & $0.15_{-0.04}^{+0.04}$  & 0.95  & $0.18_{-0.05}^{+0.06}$ & 0.96 \\ \vspace{0.15cm}
                               &      &                         &                        &      & $\phi^*_1$  & $0.13_{-0.03}^{+0.04}$ & 1.45 & $0.16_{-0.04}^{+0.04}$  & 1.47  & $0.10_{-0.06}^{+0.04}$ & 1.50 \\ \vspace{0.15cm}
                               &      &                         &                        &      & $\phi^*_L$  & $0.13_{-0.03}^{+0.03}$ & 1.58 & $0.17_{-0.04}^{+0.05}$  & 1.65  & $0.07_{-0.03}^{+0.05}$ & 1.78 \\ \vspace{0.15cm}
                               &      &                         &                        &      & $\phi^*_w$  & $0.16_{-0.03}^{+0.03}$ & 1.12 & $0.18_{-0.04}^{+0.04}$  & 1.16  & $0.10_{-0.05}^{+0.05}$ & 1.26 \\ \vspace{0.15cm}
                               &      &                         &                        &      & control     & $0.03_{-0.02}^{+0.03}$ & 2.29 & $0.03_{-0.02}^{+0.03}$  & 2.30  & $0.04_{-0.03}^{+0.03}$ & 2.33 \\ \vspace{0.15cm}
$z_c < 0.313$          & 1138 &  $2.16_{-0.12}^{+0.12}$ & $0.78_{-0.04}^{+0.05}$ & 1.58 & $\phi_1$    & $0.16_{-0.04}^{+0.04}$ & 1.58 & $0.09_{-0.04}^{+0.04}$  & 1.74  & $0.28_{-0.05}^{+0.06}$ & 2.37 \\ \vspace{0.15cm}
                               &      &                         &                        &      & $\phi_L$    & $0.16_{-0.03}^{+0.04}$ & 1.88 & $0.12_{-0.05}^{+0.04}$  & 1.96  & $0.27_{-0.07}^{+0.05}$ & 2.40 \\ \vspace{0.15cm}
                               &      &                         &                        &      & $\phi_w$    & $0.17_{-0.03}^{+0.03}$ & 0.94 & $0.14_{-0.04}^{+0.04}$  & 0.98  & $0.23_{-0.06}^{+0.06}$ & 1.09 \\ \vspace{0.15cm}
                               &      &                         &                        &      & $\phi^*_1$  & $0.16_{-0.03}^{+0.03}$ & 1.07 & $0.16_{-0.04}^{+0.04}$  & 1.07  & $0.16_{-0.05}^{+0.05}$ & 1.07 \\ \vspace{0.15cm}
                               &      &                         &                        &      & $\phi^*_L$  & $0.14_{-0.03}^{+0.04}$ & 1.46 & $0.13_{-0.04}^{+0.04}$  & 1.46  & $0.17_{-0.06}^{+0.06}$ & 1.48 \\ \vspace{0.15cm}
                               &      &                         &                        &      & $\phi^*_w$  & $0.17_{-0.03}^{+0.03}$ & 0.93 & $0.16_{-0.04}^{+0.04}$  & 0.93  & $0.16_{-0.06}^{+0.06}$ & 0.93 \\ \vspace{0.15cm}
                               &      &                         &                        &      & control     & $0.02_{-0.01}^{+0.02}$ & 1.26 & $0.04_{-0.02}^{+0.03}$  & 1.33  & $0.02_{-0.01}^{+0.02}$ & 1.25 \\ \vspace{0.15cm}
$z_c \geqslant 0.313$  & 1137 &  $1.80_{-0.16}^{+0.17}$ & $0.79_{-0.07}^{+0.07}$ & 0.74 & $\phi_1$    & $0.20_{-0.05}^{+0.05}$ & 0.83 & $0.25_{-0.07}^{+0.06}$  & 0.90  & $0.13_{-0.08}^{+0.08}$ & 0.91 \\ \vspace{0.15cm}
                               &      &                         &                        &      & $\phi_L$    & $0.23_{-0.05}^{+0.05}$ & 1.32 & $0.28_{-0.08}^{+0.06}$  & 1.36  & $0.17_{-0.08}^{+0.10}$ & 1.39 \\ \vspace{0.15cm}
                               &      &                         &                        &      & $\phi_w$    & $0.20_{-0.05}^{+0.06}$ & 0.84 & $0.25_{-0.08}^{+0.07}$  & 0.87  & $0.16_{-0.09}^{+0.10}$ & 0.89 \\ \vspace{0.15cm}
                               &      &                         &                        &      & $\phi^*_1$  & $0.24_{-0.05}^{+0.06}$ & 0.92 & $0.31_{-0.06}^{+0.07}$  & 1.03  & $0.13_{-0.08}^{+0.09}$ & 1.13 \\ \vspace{0.15cm}
                               &      &                         &                        &      & $\phi^*_L$  & $0.27_{-0.06}^{+0.06}$ & 1.10 & $0.31_{-0.07}^{+0.07}$  & 1.15  & $0.16_{-0.08}^{+0.09}$ & 1.27 \\ \vspace{0.15cm}
                               &      &                         &                        &      & $\phi^*_w$  & $0.23_{-0.05}^{+0.06}$ & 0.80 & $0.29_{-0.07}^{+0.06}$  & 0.88  & $0.10_{-0.05}^{+0.11}$ & 1.08 \\ \vspace{0.15cm}
                               &      &                         &                        &      & control     & $0.02_{-0.02}^{+0.03}$ & 1.01 & $0.02_{-0.02}^{+0.03}$  & 1.01  & $0.09_{-0.06}^{+0.07}$ & 1.25 \\                                
\hline         
\end{tabular}
\medskip
\begin{flushleft}
\textbf{Notes.} Columns: (1) Selection criteria; (2) number of clusters considered in the stack; (3), (4) and (5) results from the monopole fit; (6) Orientation angle proxy to obtain the quadrupoles; (7 - 8), (9 - 10) and (11 - 12) constrained ellipticity and reduced chi-square values from fitting both quadrupole components profiles over  $100\,h^{-1}_{70}$\,kpc$ < r < 5\,h^{-1}_{70}$\,Mpc, $100\,h^{-1}_{70}$\,kpc$ < r < 700\,h^{-1}_{70}$\,kpc and $700\,h^{-1}_{70}$\,kpc$ < r < 5\,h^{-1}_{70}$\,Mpc, respectively.
\end{flushleft}
\label{table:results}
\end{table*}

\section{Sources of bias in the ellipticity estimates}
\label{sec:bias}

In this section we discuss the possible effects that could bias our average aligned ellipticity component estimates and their impact in order to interpret our results. We are going to discuss the impact on the derived ellipticities of the misalignment between the major axis of the total mass distribution and the orientation angle estimated based on the satellite distribution. We also discuss how the selection effects of optical selected clusters can affect our ellipticity estimates. We do not intend to correct our measurements for the mentioned biases, but it is important to take them into account in order to interpret our results. All the discussed effects will bias our measurements to lower values, thus, we expect that the true projected ellipticity of the total mass distribution to be higher than the $\epsilon$ estimates. A detailed joint analysis using simulated data will be of a major importance to properly quantify these biases and will allow to link the estimates based on weak-lensing studies and the predicted projected ellipticities of dark matter halos. 

\subsection{Misalignment effect}

One of the largest source of bias is the fact that in principle, we do not know the actual major semi-axis orientation of the total mass distribution, since this is not necessary aligned with the satellite distribution. Moreover, even assuming that the satellites perfectly trace the dark mass distribution, there are many biases introduced when estimating the orientation angle according to the position of the galaxies classified as members. The main bias known as the Poisson sampling effect, is introduced due to the finite number of galaxies used to estimate the angle and it is specially important when a low number of satellites are considered or when the halo is more spherical \citep{Uitert2017}. Mentioned misalignment will result in an underestimated ellipticity measurement, which will be related to the true projected mass ellipticity through Eq.\ref{eq:ellip}.

If it is assumed that the galaxy distribution properly trace the dark matter, the ellipticities can be corrected by the Poisson sampling effect using simulations to evaluate the misalignment introduced as a function of the number of satellites and to estimate the correction factor. \citet{Shin2018} concluded that for the redMaPPer clusters the estimated ellipticity has to be corrected by a factor of $\sim 1.33$, which represents a misalignment of $18^{\circ}$. The other sources of uncertainty introduced when computing the orientation angle are an edge effect, since members are selected within a circular aperture, and an effect introduced by the inclusion of interlopers. In principle, if we assume that interlopers have a random angular distribution, both effects will not introduce a bias in the estimated angle. Nevertheless these will introduce larger uncertainties in the angle estimates biasing the ellipticity to lower values.

\subsection{Halo selection projection effects}

Although redMaPPer clusters are considered one of the most homogeneous and well-calibrated catalogue of optical clusters, many studies have reported projection effects that can affect the sample \citep[e.g. ][]{Dietrich2014, Costanzi2019, Sunayama2020}. Mainly these effects have been studied in order to calibrate the richness-mass relation and to characterize their impact on cluster cosmology analyses. One of the sources of biases due to projection effects is related to the triaxiality, since optical selected clusters tend to be elongated along the line-of-sight (LOS). This effect is proven to cause overestimated lensing masses in about $3 - 6 \%$ and is more important for low-richness clusters \citep{Dietrich2014}. In that case, the lensing projected ellipticity will be also biased to lower values.

Another source of bias introduced by projection effects is related to the presence of LOS halos, which are expected to be specially significant in rich galaxy clusters due to the abundance of correlated structures around these  systems \citep{Costanzi2019}. Also, \citet{Sunayama2020} find a selection bias of optical cluster finders for clusters embedded within filaments aligned with the LOS. This increases both the observed cluster richness and the recovered lensing mass. Although the impact of LOS halos in lensing measurements have already been extensively discussed \citep{DES2020},  most studies have mainly focused on the bias introduced on the mass estimates and the mass-richness relation. We stress here the impact on measurements of cluster ellipticities since we expect that this effect may be important and induce $\epsilon$ determinations to systematically lower values. Firstly, the inclusion of interlopers in the satellite sample affects the mass distribution orientation estimate. Besides, LOS halos with different projected orientations and at different projected distances from the cluster centre, will decrease the observed projected ellipticity. Finally, these projection effects can boost the halo mass, $M_{200}$, which will affect the quadrupole fit.

\section{Discussion}
\label{sec:discuss}

In this section we discuss the results obtained considering the potential biases presented in the last section. We first discuss the projected ellipticity of the dark matter halos taking into account the average aligned ellipticity components obtained by fitting the quadrupole profiles in the inner regions. Then, we discuss how the derived projected ellipticities at the cluster outskirts might trace the accretion direction of the clusters. 

\subsection{Halo projected ellipticity}

In order to get an insight regarding the projected shape of the halos, we consider as the most representative parameter of the halo projected ellipticity, our estimates derived from fitting the quadrupole profile components up to $700h^{-1}_{70}$kpc, obtained by aligning the clusters according to the $\phi^*_1$ proxy. This selection was made considering that the surface density at the inner regions are better traced by the galaxies with a higher membership probability (see Sec. \ref{sec:results}) and since the derived ellipticities using a uniform weight are systematically larger for most of the considered samples. Nevertheless, we recall that no significant differences are observed for the different adopted weights. Taking this into account we obtain $\epsilon = 0.21 \pm 0.04$. Projected average halo ellipticity of redMaPPer clusters were previously estimated by \citet{Clampitt2016} and \citet{Shin2018}. \citet{Shin2018} estimated a mean projected halo ellipticity of $0.28 \pm 0.07$ considering a weak-lensing analysis of redMaPPer cluster within a similar redshift and richness range as those adopted in this work (0.1 < $z$ < 0.41, $20 < \lambda < 200$) but using shear catalogues based on SDSS observations. If we consider the correction factor for the Poisson sampling effect estimated by these authors, our corrected ellipticity measurement, $0.28 \pm 0.05$, is in excellent agreement with their estimate. This result is in agreement with the expectation from $\Lambda$CDM numerical simulations \citep{Despali2017}. However it is important to take into account that we only consider for the comparison the introduced bias by the Poisson sampling effect. As discussed in the previous section, the other sources of misalignment and the selection effect of photometric clusters, will result in an underestimated projected ellipticity. 

We obtain larger $\epsilon$ values for the samples selected at high-redshift ($z \geq 0.313$) and low-mass clusters. For the low-mass clusters we estimate an average aligned ellipticity component of $0.27 \pm 0.07$ while for the high-mass clusters we obtain $0.16 \pm 0.04$. On the other hand, we obtain $0.16 \pm 0.04$ and $0.31^{+0.07}_{-0.06}$ for the low- and high-redshift samples. In order to evaluate the error introduced by the sample dispersion, we performed a randomization test by fitting the parameters from 100 monopole and quadrupole profiles, derived by randomly selecting half of the total sample of clusters. The derived $\epsilon$ distribution has a mean of 0.21 and a standard deviation of 0.04, in excellent agreement with the values obtained for the total sample. Taking into account the previous analysis, we conclude that observed differences are not produced by sampling effects. 

According to $\Lambda$CDM numerical simulations, it is expected that galaxy clusters at higher redshifts to be less spherical, since they are more affected by the direction of their major last merger, which is in agreement with our results. Also, the observed tendency between the low- and high-redshift cluster samples is obtained regardless which weight is considered in order to compute the cluster orientation angle. Furthermore, the average projected ellipticity estimated for the high-redshift sample ($\langle z \rangle = 0.35$), is in agreement with the analysis of \citet{Okabe2020} (see Fig. 9 of their paper) based on the cosmological hydrodynamical simulation Horizon-AGN \citep{Dubois2014}.

Although with a low significance, the result obtained for the low- and high-mass samples is in disagreement with the expectation from isolated halos in $\Lambda$CDM numerical simulations. Given that more massive halos are formed later, they are expected to be less spherical. To test our result, we consider a higher redshift gap for the background galaxy selection, taking  Z\_BEST $> z_c+0.3$, and we recompute the profiles. This test is motivated by the fact that a larger fraction of red galaxies is expected at the location of higher mass clusters. These cluster member objects can contaminate the sample of background galaxies inducing a dilution of the lensing signal. Derived ellipticities from the profiles obtained with this new background galaxy sample are $0.11\pm0.06$ and $0.31\pm0.08$ for the high- and low-mass cluster samples, respectively. This suggests that the observed gap is not due to a bias introduced by the selection of background galaxies. 

It is important to highlight that the projected ellipticities estimates for the low- and high-mass samples, are in agreement within $1\sigma$. Moreover, if we consider a different weight to estimate the orientation angle of the mass distribution than the adopted uniform weight, i.e. if the aligned ellipticity component is obtained according to $\phi^*_d$ or $\phi^*_L$, the difference between the fitted values is even lower. However, we cannot neglect the possibility that the effects introduced by the contribution of the surrounding halos on the LOS can bias the projected ellipticity estimate to lower values, affecting mainly the higher-richness sample.  The use of numerical simulations to measure the projected ellipticity, mimicking the same approach of this analysis (i.e. defining an axis from the galaxy distributions and measuring the lensing signal accounting for the matter along the LOS) can be of great importance to study the biases discussed and to assess quantitatively their impact on the lensing estimates.


\subsection{Projected density distribution at the cluster outskirts}

Here we intend to discuss how the projected density mass is distributed at the outskirts of the clusters by considering the fitted projected ellipticities at the outer regions of the quadrupole profiles ($r > 700h^{-1}_{70}$kpc). In order to do that, we consider the projected ellipticities derived by aligning the clusters taking into account $\phi_1$, since, as discused in Sec. \ref{sec:results}, the whole sample of members traces better the mass distribution at larger distances from the cluster centre (see Fig. \ref{fig:emedians} and \ref{fig:erranges}). The membership cut can be useful to discard interlopers, since their inclusion could result in a wrongly estimate of the orientation angle, thus underestimating the projected ellipticity. However, this sample of satellites could be also including galaxies that have been recently accreted by the cluster. $p_{mem}$ is computed according to the  galaxy  colour, luminosity  and projected distance from the cluster centre. Therefore, satellites with $p_{mem} < 0.5$ are in principle dimmer, bluer and located at larger distances from the cluster centre. The observed result might suggest that the outer region is better traced when $p_{mem} < 0.5$ satellites are included, since this sample include bluer galaxies that were accreted later and thus follow the mass distribution at the cluster outskirts.

Similarly to the inner region, which is more related to the halo projected elongation, low-mass systems tend to show larger aligned ellipticity component values in the outer regions, being $0.28_{-0.09}^{+0.10}$, while for the high-mass sample we obtain $0.20_{-0.06}^{+0.04}$. Nevertheless, both results are in agreement taking into account the errors. On the other hand, derived ellipticities at the outskirts for the low-redshift  sample is larger than for high-redshift clusters, being $0.28_{-0.05}^{+0.06}$ and $0.13 \pm 0.08$, respectively. This tendency can be also observed in the right panel of Fig. \ref{fig:erranges}, in which the inner and outer ellipticities are uncorrelated for the two redshifts samples, being the inner regions less spherical for the low-redshift sample, while the high-redshift sample is less spherical at the outskirts. This result can be indicating that the projected density distribution is better traced by the inclusion of bluer galaxies at lower redshifts. Also, it might be related to the cosmic web evolution, since it is expected for the filamentary distribution to have higher densities and a steeper profile at lower redshifts \citep{Smith2012,Cautun2014,Kraljic2018}. This might suggest that the accretion region is better constrained at lower redshifts. 

\section{Summary and conclusions}
\label{sec:conclusion}

In this work we presented a study of the surface mass density shape of galaxy clusters, through the estimated aligned component of the projected ellipticity using weak-lensing techniques. We used the optically selected SDSS redMaPPer clusters within a redshift and a richness range of $0.1 \leq z < 0.4$ and $20 \leq \lambda < 150$, respectively. The weak-lensing analysis was performed taking advantage of the combination of four high-quality shear catalogs. We modelled the derived lensing signal taking into account a multipole expansion of the surface density distribution. The total surface density was modelled considering two terms that include the monopole and a quadrupole, where the quadrupole is proportional to the projected ellipticity. Quadropole profiles were computed by aligning the clusters taking into account the satellite distribution, for which we consider six different proxies. Finally, we fit the profiles in three projected radius ranges to obtain information regarding the shape of the mass distribution at the inner and outer regions of the clusters. 

Derived ellipticities by fitting the profile up to $700h^{-1}_{70}$kpc are larger if a membership cut is applied to the satellites considered to estimate the orientation angle. Therefore, the inner regions are better traced by the galaxies that have a higher membership probability. On the other hand, projected ellipticity values are larger if the profiles are fitted in the outer regions ($> 700h^{-1}_{70}$kpc) when all the galaxies classified as members are included to estimate the orientation of the mass distribution. Therefore, the inclusion of bluer galaxies mainly located at larger projected radius, trace better the density distribution at the outskirt of the clusters. 

To study the mean halo projected ellipticity we consider the estimates derived by fitting the profile at the inner regions. For the total sample of clusters we derive a projected ellipticity of $0.21 \pm 0.04$. If we consider the Poisson sampling effect, which takes into account that the orientation angle is estimated from a finite number of galaxies, we obtain $0.28 \pm 0.04$, in excellent agreement with previous estimates \citep{Shin2018}. Although this result is also in agreement with the predicted by numerical simulations \citep{Despali2017}, a detailed quantification of possible biases is important in order to link the weak-lensing projected ellipticity estimates with the dark matter halo expected elongation. 

We also analyse the dependence of the derived cluster projected ellipticity estimates and the mean mass and redshift of these systems. This is accomplished by splitting the sample into two bins considering the median values of richness and redshift of the total sample, respectively. We obtain a larger mean projected ellipticity for the high-redshift cluster, in agreement with the expectation of $\Lambda$CDM numerical simulations. Also, we obtain a tendency for a smaller projected ellipticity in the high-mass cluster sample, although its significance is low. Moreover, it should be considered that a higher contribution of the surrounding clusters distributed mostly along the line-of-sight, could be biasing the projected ellipticity of the richest clusters to lower values. A further analysis on the projection effects of optically-selected clusters and their influence on the lensing ellipticity measurements is required to characterize properly the mean projected halo elongation.

Finally, we study the density mass distribution at the outskirts, considering the ellipticity estimates obtained by fitting the profile at the outer regions and the whole sample of galaxy members to align the clusters. No significant differences are observed for the low- and high-mass cluster samples. Nevertheless, we derive a larger mean ellipticity for the low-redshift cluster sample than for the high-redshift sample. This result can be indicating a more constrained direction of the matter accretion region at lower redshifts. 

Taking advantage of the combination of high-quality weak-lensing data, we presented the analysis of the projected shape of the cluster-size halos and its relation with the mass and redshift. We study the mass distribution at the cluster outskirts, which can be related to the large-scale structure and can provide information about the matter accretion of these galaxy systems. This work anticipates the full potential of the weak gravitational lensing techniques in the study of the average projected mass distribution shape of galaxy clusters. Recently release of DES-Y1 based on the Dark Energy Survey \citep[DES,][]{DES}, provides a new sample of redMaPPer clusters spanning a larger redshift and richness range that will allow a deeper study of halo shape and the general projected mass distribution of galaxy clusters. Moreover, new upcoming data based on DES and on the forthcoming 
`The Rubin Observatory Legacy Survey of Space and Time' \citep[LSST,][]{LSST} and on the EUCLID mission \citep{Amendola2013}, will offer a larger statistical sample of galaxy clusters plus high-quality shear catalogues that will allow a deeper study on the relation of the projected halo elongation with the redshift and mass. These studies combined with the predictions from simulations, can shed light regarding the shape evolution of dark matter halos and their accretion history.

\section*{Acknowledgements}

This work was partially supported by the Consejo Nacional
de  Investigaciones  Cient\'ificas  y  T\'ecnicas  (CONICET, Argentina), the Secretar\'ia de Ciencia y Tecnolog\'ia de la Universidad Nacional de C\'ordoba (SeCyT-UNC, Argentina), the Brazilian Council for Scientific and Technological Development (CNPq) and the Rio de Janeiro Research Foundation (FAPERJ). 
We acknowledge the PCI BEV fellowship program from
MCTI and CBPF.
MM acknowledges FAPERJ and CNPq for financial support. FORA BOZO.
HYS acknowledges the support from NSFC of China under grant 11973070, the Shanghai Committee of Science and Technology grant No.19ZR1466600 and Key Research Program of Frontier Sciences, CAS, Grant No. ZDBS-LY-7013.
LVW acknowledges the support from the Natural Sciences and Engineering Research Council of Canada.
This  work  is  based  on  observations  obtained  with
MegaPrime/MegaCam,  a  joint  project  of  CFHT  and
CEA/DAPNIA, at the Canada--France--Hawaii Telescope
(CFHT), which is operated by the National Research
Council (NRC) of Canada, the Institut National des Sciences de l'Univers of the Centre National de la Recherche Scientifique (CNRS) of France, and the University of
Hawaii. The Brazilian partnership on CFHT is managed
by the Laborat\'orio Nacional de Astrof \'isica (LNA). We
thank the support of the Laborat\'orio Interinstitucional de
e-Astronomia (LIneA). We thank the CFHTLenS team
for their pipeline development and verification upon which
much of the CS82 survey pipeline was built.\\
This research used the facilities of the
Canadian Astronomy Data Centre operated by the National Research
Council of Canada with the support of the Canadian Space Agency.
RCSLenS data processing was made possible thanks to significant
computing support from the NSERC Research Tools and Instruments grant
program.
Based on data products from observations made with ESO Telescopes at the La Silla Paranal Observatory under programme IDs 177.A-3016, 177.A-3017 and 177.A-3018. 





\bibliographystyle{mnras}
\bibliography{references}


\appendix

\section{Testing the combination of the weak-lensing catalogs}
\label{app:test}
\begin{figure}
    \centering
    \includegraphics[scale=0.6]{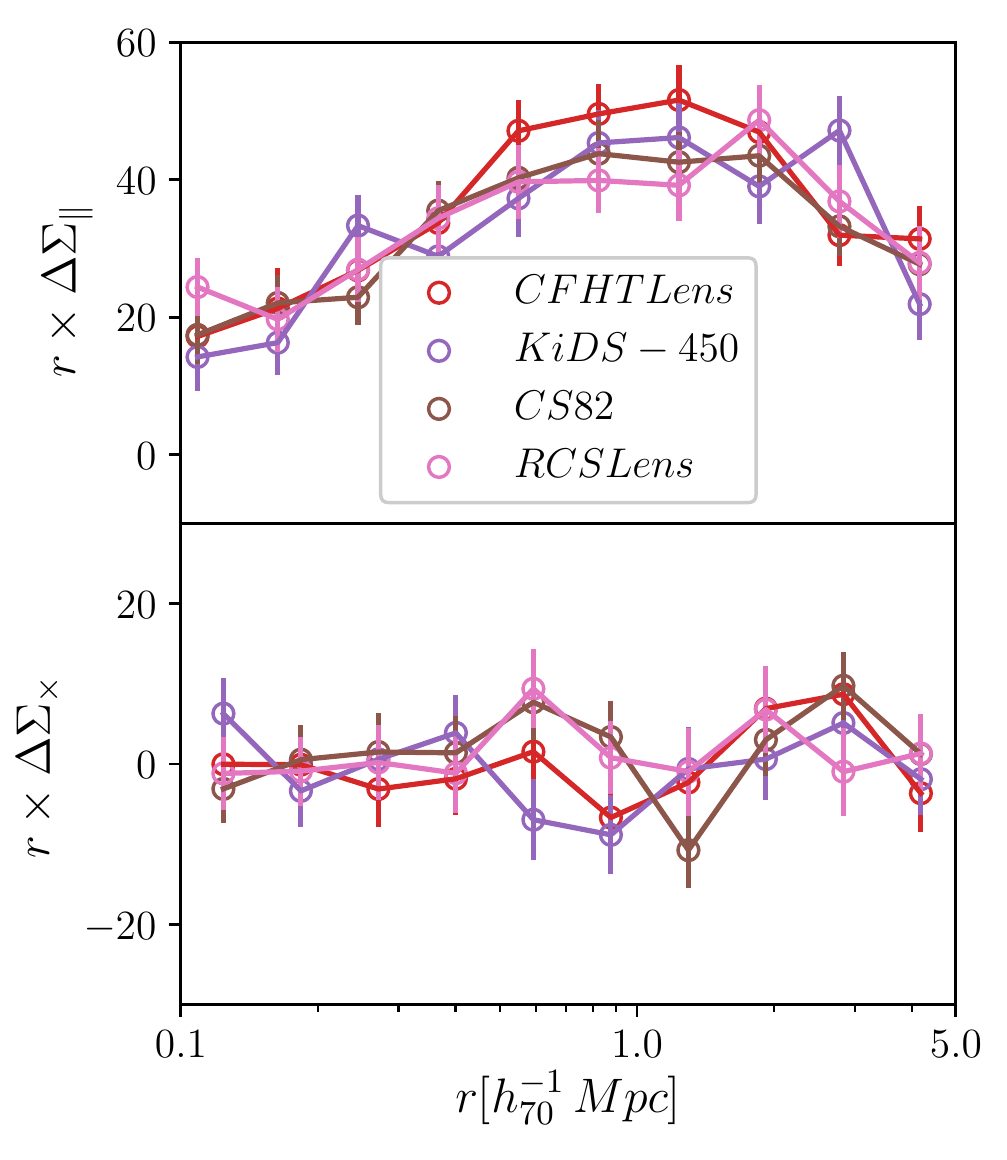}
    \caption{Tangential (upper panel) and cross (bottom panel) component of the shear monopole scaled according to the projected distance in units of [Mpc\,$M_\odot$\,pc$^{-2}$], computed considering the used shear catalogues individually}
    \label{fig:monotest}
\end{figure}

\begin{figure*}
    \centering
    \includegraphics[scale=0.6]{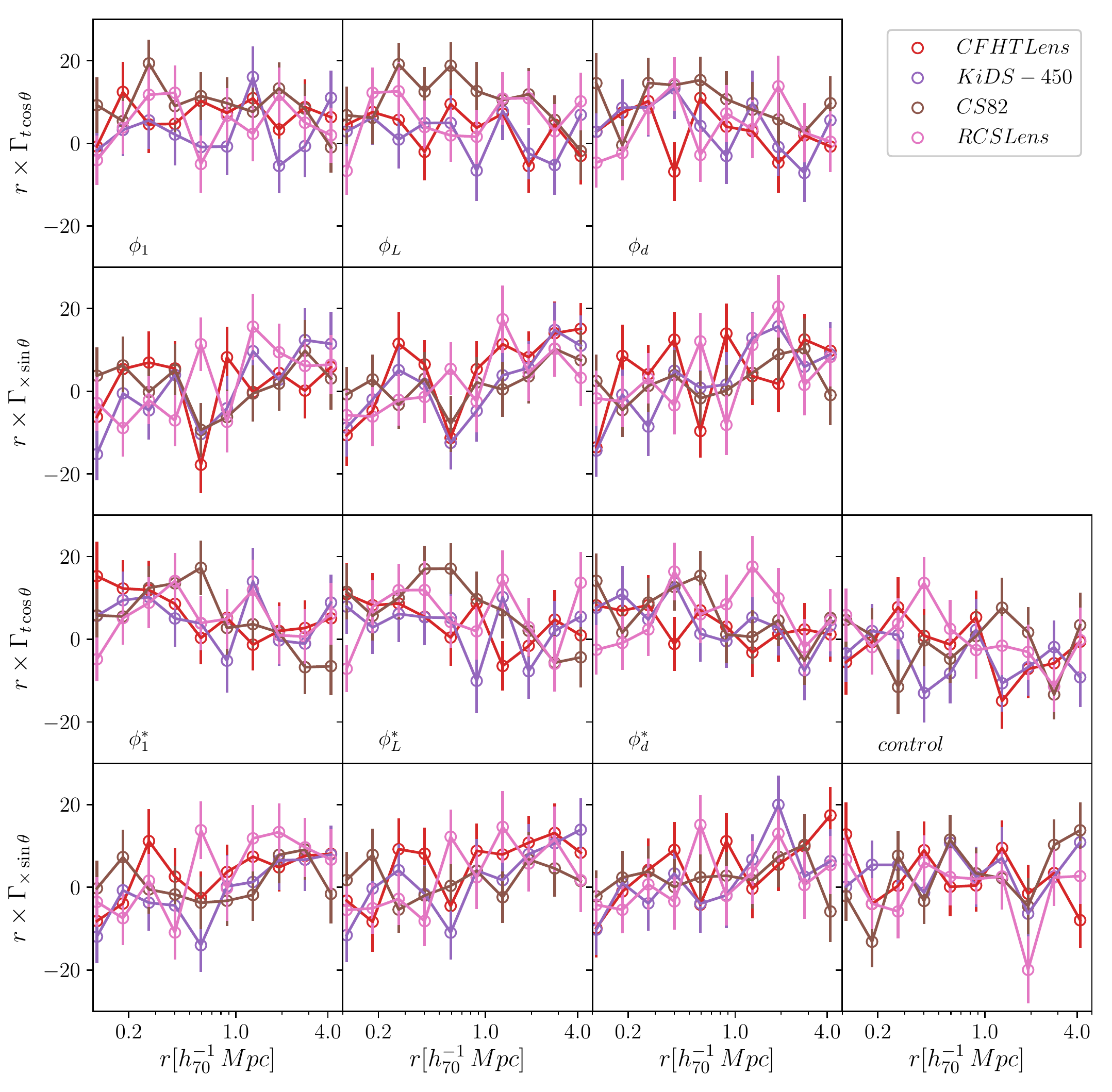}
    \caption{Tangential ($\Gamma_{\rm{t} \cos{2\theta}}$) and cross ($\Gamma_{\rm{\times} \sin{2\theta}}$) component of the quadrupoles scaled according to the projected distance in units of [Mpc\,$M_\odot$\,pc$^{-2}$], computed considering the used shear catalogues individually for the different orientation angle proxies.}
    \label{fig:qtest}
\end{figure*}

In order to test the combination of the weak-lensing shear catalogues presented in Sec. \ref{sec:data}, we compute the monopole and quadruole profile components considering the individual shear catalogs. With this purpose we consider the total sample of clusters used in this work. Derived monopole and quadrupole profiles scaled according to the projected distance to emphasize the differences are shown in Fig. \ref{fig:monotest} and Fig. \ref{fig:qtest}, respectively. As it can be noticed, not significant biases are observed in the computed profiles. We also fitted the monopoles computed using the shear catalogues separately and the quadrupoles computed considering the orientation angle estimated using the whole sample of galaxy members and a uniform weight. Results are presented in Table \ref{table:indcat}. All the fitted parameters are in agreement within $1\sigma$ with the resultant fitted values derived using the combined catalogs. We also show the contribution to the total sample of background galaxies for each catalog, being the CFHTLens catalogue the most prevailing shear catalog.

\begin{table}
\caption{Fitted parameters from the monopole and quadrupole profile components computed using the combined weak-lensing shear catalogues and each individual catalog.}
\begin{tabular}{c c c c c c}
\hline
\hline
\rule{0pt}{1.05em}%
Shear       & $N_L$      &      $N_S/N_T$     &  $M_{200}$ & $p_{cc}$ & $\epsilon$  \\
  catalogue                   &            &         &                       &\\ 
\hline
\rule{0pt}{1.05em}   
\vspace{0.15cm}
Combined & 2275 & 1 & $2.06_{-0.11}^{+0.09}$ & $0.79_{-0.04}^{+0.04}$ & $0.17_{-0.03}^{+0.03}$ \\ 
\hline
\rule{0pt}{1.05em}   
\vspace{0.15cm}
CFHTLens & 311 & 0.30 & $2.09_{-0.15}^{+0.18}$ & $0.85_{-0.08}^{+0.07}$ & $0.18_{-0.05}^{+0.06}$ \\ \vspace{0.15cm}
CS82     & 662 & 0.24 & $1.84_{-0.13}^{+0.15}$ & $0.82_{-0.07}^{+0.06}$ & $0.25_{-0.05}^{+0.05}$ \\ \vspace{0.15cm}
KiDS-450 & 684 & 0.28 & $2.06_{-0.20}^{+0.16}$ & $0.74_{-0.07}^{+0.07}$ & $0.09_{-0.05}^{+0.06}$ \\ \vspace{0.15cm}
RCSLens  & 1005 & 0.19 & $1.93_{-0.18}^{+0.18}$ & $0.81_{-0.07}^{+0.07}$ & $0.15_{-0.06}^{+0.05}$ \\
\hline         
\end{tabular}
\medskip
\begin{flushleft}
\textbf{Notes.} Columns: (1) Weak-lensing shear catalogs; (2) number of clusters considered for the stacking, (3) Fraction of background galaxies included in each catalog; (4) mass in units of [$h^{-1}_{70} 10^{14} M_\odot$] and (5) fraction of well-centred clusters derived from the monopole fit; (6) constrained projected ellipticity.
\end{flushleft}
\label{table:indcat}
\end{table}

\section{Quadrupole profiles}
\label{app:quadprofiles}

\begin{figure*}
    \centering
    \includegraphics[scale=0.65]{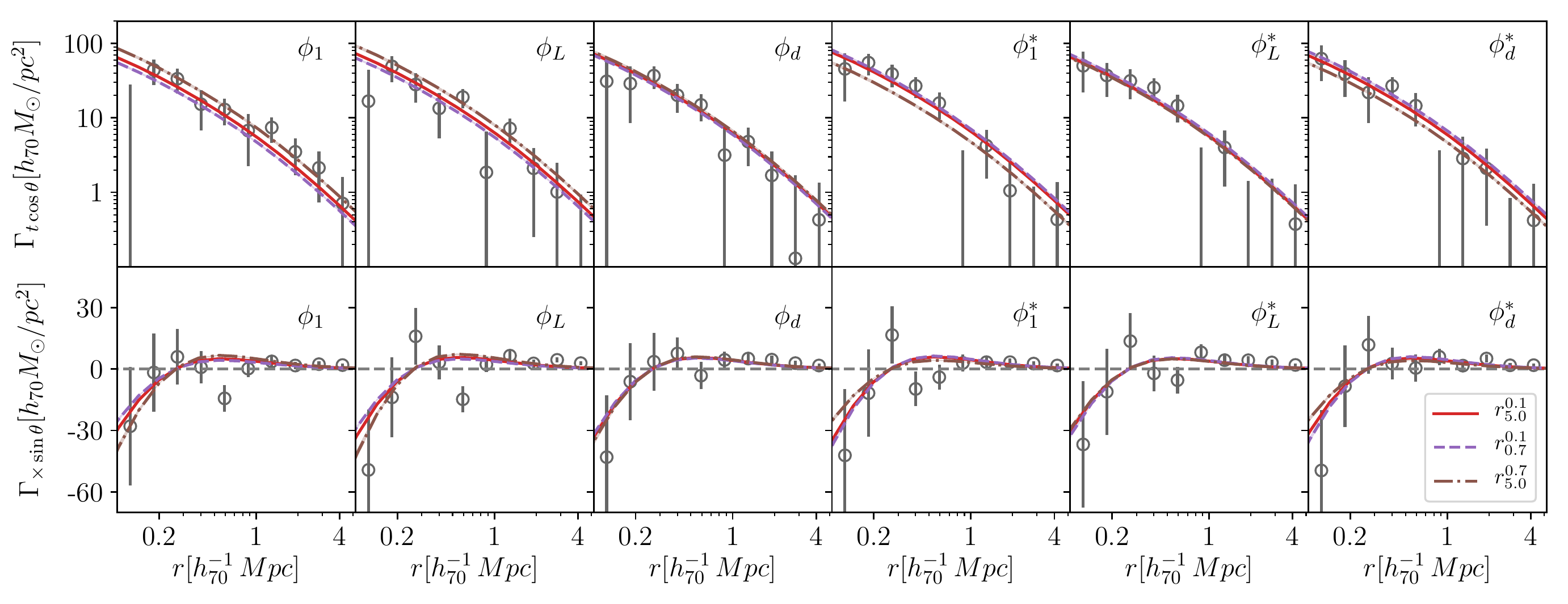}
    \caption{Derived quadrupole tangential and cross component profiles, $\Gamma_{t \cos{\theta}}$ and  $\Gamma_{\times \sin{\theta}}$, for the total sample of redMaPPer clusters analysed in this work. Best fits for the three different considered radius ranges are also shown. Quadrupoles are obtained taking into account the derived orientation angles computed for each cluster according to the proxies defined in Sec. \ref{sec:orientation} ($\phi_1$,$\phi_L$,$\phi_d$,$\phi^*_1$,$\phi^*_L$ and $\phi^*_d$)}
    \label{fig:qtotal}
\end{figure*}

\begin{figure*}
    \centering
    \includegraphics[scale=0.65]{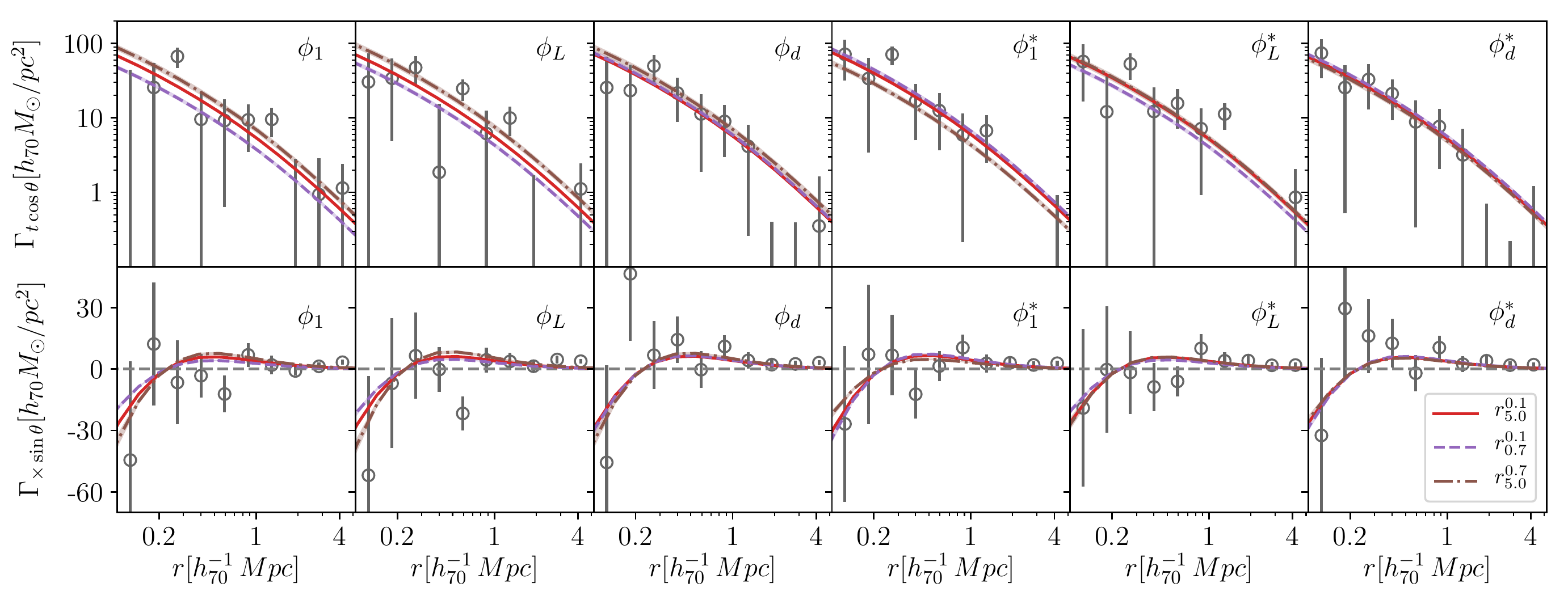}\\
    \includegraphics[scale=0.65]{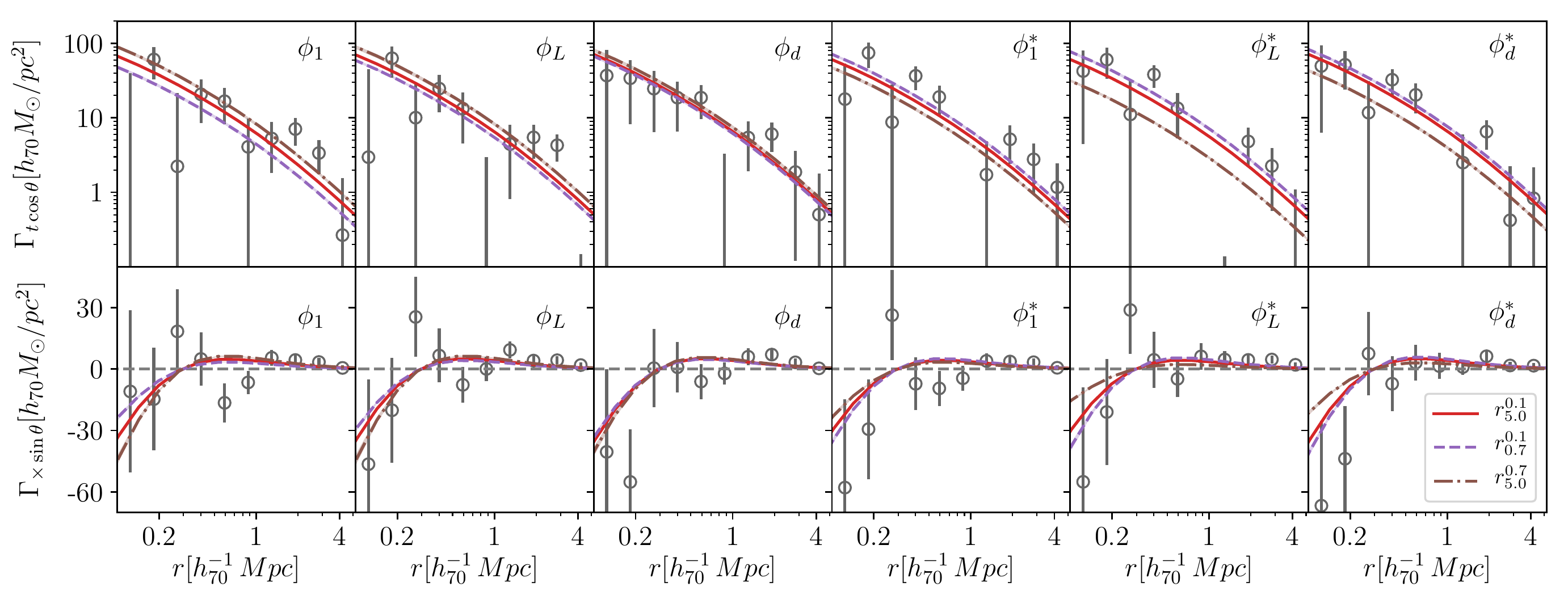}
    \caption{Idem as in Fig. \ref{fig:qtotal} but for the low-mass (upper panel) and high-mass (bottom panel) cluster samples, selected according to the median of the richness distribution.}
    \label{fig:qmbin}
\end{figure*}

\begin{figure*}
    \centering
    \includegraphics[scale=0.65]{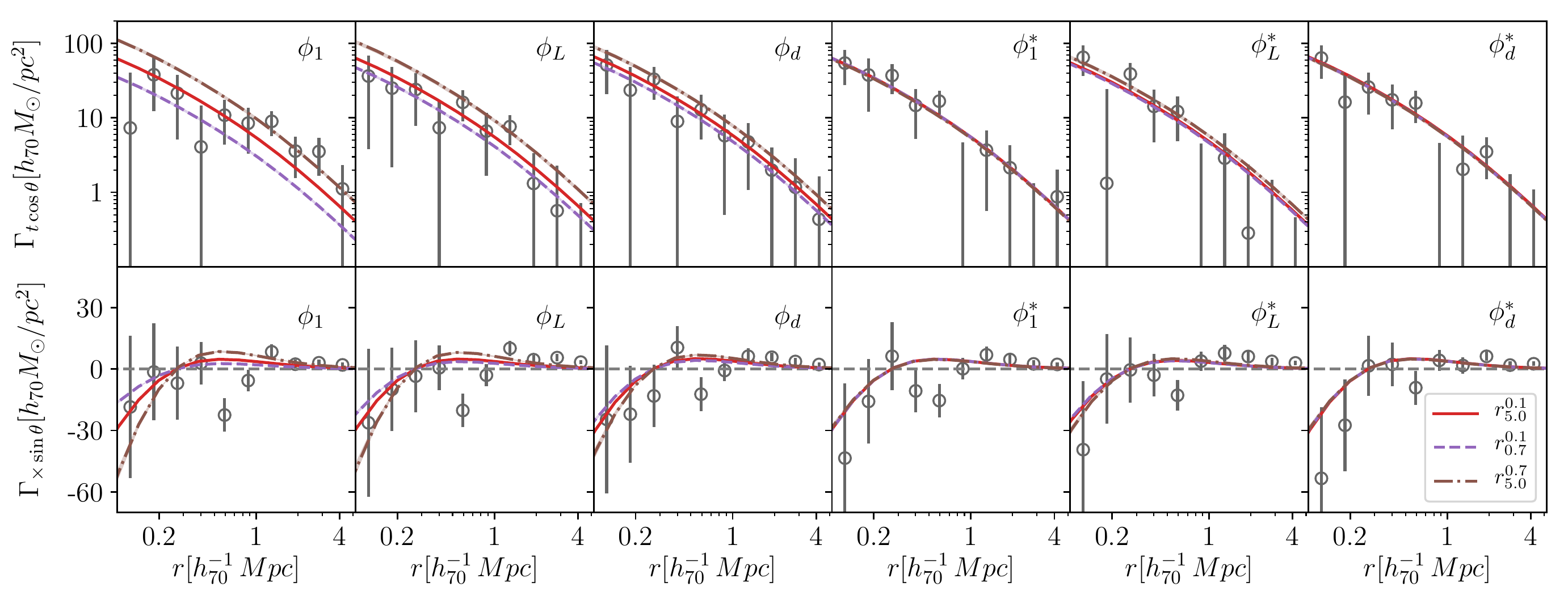}\\
    \includegraphics[scale=0.65]{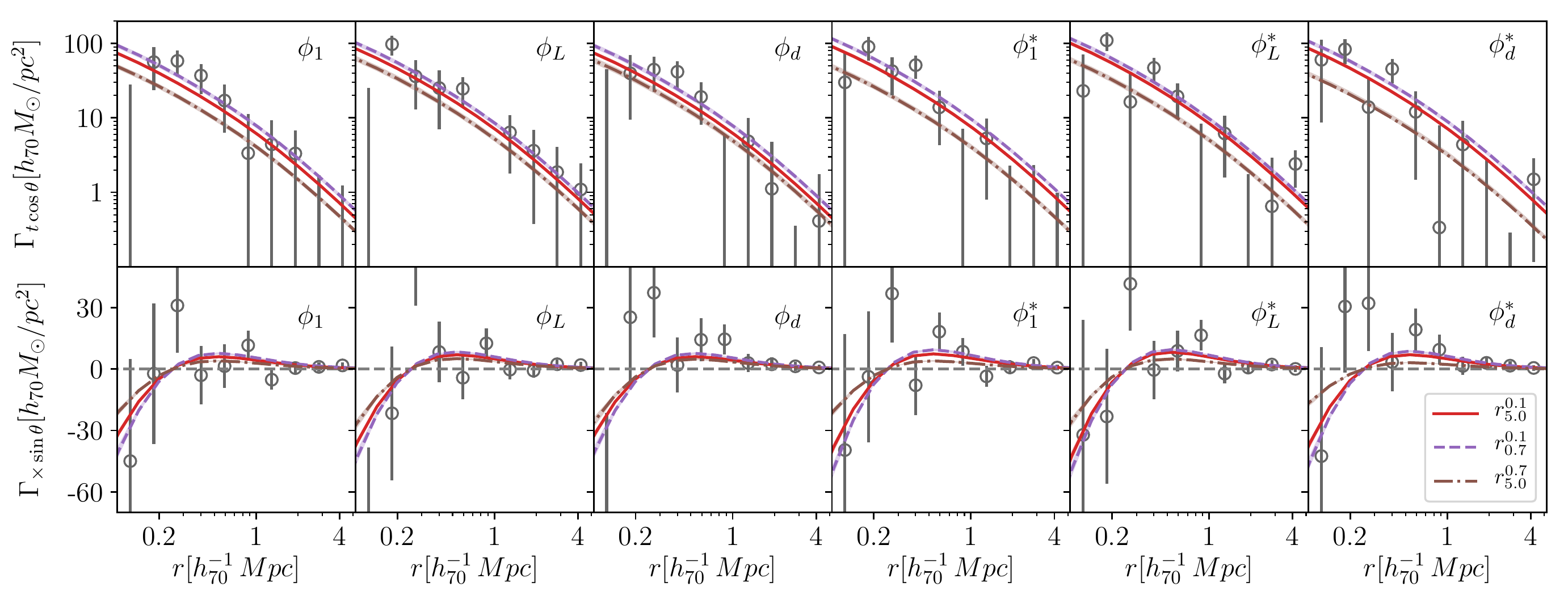}
    \caption{Idem as in Fig. \ref{fig:qtotal} but for the low-redshift (upper panel) and high-reshift (bottom panel) cluster samples, selected according to the median of the redshift distribution.}
    \label{fig:qzbin}
\end{figure*}

In this Appendix we present the quadrupole profiles together with their best fit models. We show in Fig. \ref{fig:qtotal} the quadrupoles derived for the total sample of the redMaPPer clusters. In Fig. \ref{fig:qmbin} and in Fig. \ref{fig:qzbin} we show the profiles computed for the high- and low-mass cluster samples and for the high- and low-redshift samples, respectively. Derived best fitted ellipticities are detailed in Table \ref{table:results}.

\bsp	
\label{lastpage}
\end{document}